\documentclass[twocolumn,showpacs,preprintnumbers,%
amsmath,amssymb,aps,floatfix]{revtex4}

\usepackage{graphicx}% Include figure files
\usepackage{dcolumn}% Align table columns on decimal point
\usepackage{bm}% bold math
\usepackage[FIGBOTCAP]{subfigure}

 \newcommand{\md}         {\,\textrm{d}} % differential

\begin{document}

\title{Ionization of molecular hydrogen and deuterium by a frequency-doubled 
Ti:sapphire laser pulses}  

\author{Yulian V. Vanne}
\author{Alejandro Saenz}%
\affiliation{%
AG Moderne Optik, Institut f\"ur Physik, Humboldt-Universit\"at
         zu Berlin, 
         Hausvogteiplatz 5-7, D\,--\,10\,117 Berlin, Germany}%

\date{\today}%  

\begin{abstract}
A theoretical study of the intense-field single ionization of 
molecular hydrogen or deuterium oriented either parallel or 
perpendicular to a 
linear polarized laser pulse (400\,nm) is performed for different 
internuclear separations and pulse lengths in an intensity range of 
$(2-13)\times10^{13}\,$W\,cm$^{-2}$. The investigation is based on 
a non-perturbative treatment that solves the full 
time-dependent Schr\"odinger equation of both correlated electrons within the 
fixed-nuclei and the dipole approximation. The results for 
various internuclear separations are used to obtain the ionization yields  
of molecular hydrogen and deuterium in their ground vibrational states.
An atomic model is used to identify the influence of the intrinsic diatomic
two-center character of the problem.
 
\end{abstract}


\maketitle

%%%%%%%%%%%%%%%%%%%%%%%%%%%%%%%%%%%%%%%%%%%%%%%%%%%%%%%%%%%%%%%%%%%%%%%%%%%%%
%%%%%%%%%%%%%%%%%%%%%%%%%%%%%%%%%%%%%%%%%%%%%%%%%%%%%%%%%%%%%%%%%%%%%%%%%%%%%
\section{\label{sec:Intro}Introduction} 
%%%%%%%%%%%%%%%%%%%%%%%%%%%%%%%%%%%%%%%%%%%%%%%%%%%%%%%%%%%%%%%%%%%%%%%%%%%%%
%%%%%%%%%%%%%%%%%%%%%%%%%%%%%%%%%%%%%%%%%%%%%%%%%%%%%%%%%%%%%%%%%%%%%%%%%%%%%

The theoretical and experimental investigation of the interaction of molecules
with intense laser fields remains one of the most challenging problems of 
atomic, molecular, and optical physics. Despite the numerous experimental work 
on molecules in intense ultrashort laser pulses during the last decades 
(see, e.\,g., \cite{sfm:post04,sfm:hert06} for reviews) a full understanding 
of the 
influence of the molecular structure on the strong-field response is still 
lacking. However, such an understanding is a prerequisite for the recently 
proposed techniques that aim for the time-resolved imaging of changes of the 
electronic structure during chemical reactions. The development of such
imaging techniques is driven by successful pioneering experiments in which the 
strong-field induced high-harmonic radiation \cite{sfm:itat04} or the in the 
ionization process ejected electrons \cite{sfm:meck08} were used to image the
highest-occupied molecular orbitals of molecular nitrogen or oxygen.  

In order to achieve a three-dimensional image of the electron density or even 
an electronic orbital it is of course important to perform a spatially 
resolved measurement. Often, field-free alignment by means of rotational 
wave-packets \cite{sfm:lars99b,sfm:stap03} is adopted for providing  
angle-resolved molecular strong-field data 
\cite{sfm:litv03,sfm:zeid06,sfm:pavi07,sfm:meck08}, but in 
\cite{sfm:stau09,sfm:magr09} alternative techniques without alignment 
are used. 
These experimental efforts are accompanied by a number of theoretical 
investigations of the orientational dependent ionization probability 
of molecules in intense laser pulses. Since most molecular systems 
require some approximative treatment like the strong-field approximation 
based additionally on the single-active electron approximation (SAE) 
\cite{sfm:beck04,sfm:kjel06,sfm:milo06,sfm:usac06} or an effective 
independent-particle model like the time-dependent density-functional 
theory (TD-DFT) \cite{sfm:teln09}, the simplest neutral stable 
molecule, H$_2$ whose orientational dependence
\cite{sfm:boch08,sfm:stau09,sfm:magr09} or angular distribution of 
ejected electrons \cite{sfm:wilb08} was recently investigated 
experimentally is, in principle, an 
attractive alternative. Solutions of the time-dependent Schr\"odinger 
equation (TDSE) describing both electrons of H$_2$ exposed to an 
intense laser pulse in full dimensionality became recently 
available \cite{sfm:haru00,sfm:awas05,sfm:pala06}. However, these 
calculations were restricted to a parallel orientation of the molecule 
with respect to the field axis of a linear-polarized laser. This 
simplifies the treatment drastically, since the problem reduces to 
five spatial dimensions, as the cylindrical symmetry is preserved. 

Theoretical investigations of the orientational dependence 
of the strong-field behavior of H$_2$ are thus rather limited so far. 
This includes studies within the lowest-order perturbation theory 
(LOPT) \cite{sfm:apal02}, TD-DFT \cite{sfm:uhlm06}, or a Hartree-Fock 
based SAE approach \cite{sfm:niko07}. The validity of the SAE 
(and simplified models 
like the molecular Ammosov-Delone-Krainov tunneling model (MO-ADK) 
or the molecular strong-field approximation (MO-SFA)) that reduces 
the problem to three spatial dimensions was investigated 
in \cite{sfm:awas08} and found to be problematic 
especially for few-photon processes. Only very recently the first 
investigation of the orientational dependence of the behavior of 
H$_2$ in ultrashort intense laser pulses based on the TDSE 
was presented \cite{sfm:vann08}, and thus a full treatment of 
all six spatial dimensions of the two electrons was achieved.   
In \cite{sfm:vann08} the single-ionization yield for a parallel 
and a perpendicular orientation was compared as a function of the 
wavelength, spanning an interval between about 50 and 400\,nm. 
Considering two different laser intensities and nuclear distances, 
only a brief idea of the influence of these parameters was provided. 

In the context of imaging it is, however, important to investigate  
in which way laser intensity or the quantum-mechanically unavoidable 
zero-point vibrational motion may blur the obtained image. On the other 
hand, the strong-field response itself may be used to visualize nuclear 
dynamics with sub-femtosecond resolution 
\cite{sfm:niik02,sfm:niik03,sfm:bake06,sfm:goll06,sfm:ergl06,sfm:bake08,
sfm:fang08a}. 
In fact, for molecules like H$_2$ a strong influence of nuclear 
motion on the strong-field ionization behavior was predicted on the 
basis of a simple model in \cite{sfm:saen00c} and later confirmed 
by {\it ab initio} calculations of quasistatic rates 
\cite{sfm:saen00a,sfm:saen00b,sfm:saen02a,sfm:saen02b} and 
full TDSE calculations \cite{sfm:awas06}. An experimental confirmation 
for 800\,nm radiation was achieved by the observation of strong 
deviations from the Franck-Condon distribution of the formed H$_2^+$ 
vibrational states \cite{sfm:urba04} and the occurrence of vibrational 
wavepackets in neutral H$_2$ due to a phenomenon called 
{\it Lochfra{\ss}} \cite{sfm:ergl06} 
that had been theoretically predicted in \cite{sfm:goll06}. However, 
the responsible strong dependence of the ionization yield on the 
internuclear distance was predicted for the so-called quasi-static 
regime, i.\,e.\ for low frequencies (long wavelengths) and high 
intensities. On the other hand, the perturbative results in 
\cite{sfm:apal02} indicated that the dependence on the internuclear 
distance is expected to be rather small in this so-called multiphoton 
regime (high frequency and low intensity).  

This work investigates the intensity and internuclear-distance 
dependence of the single-ionization yield of parallel or perpendicular 
aligned H$_2$ in ultrashort linear-polarized  
laser pulses with a wavelength of about 400\,nm, as they are, e.\,g., 
experimentally available from a frequency doubling (second-harmonic 
generation) of a titanium-sapphire laser source. Besides the 
experimental relevance, the chosen wavelength is also of theoretical 
interest, since one expects six-photon ionization processes to dominate 
which lie somehow in the middle between few-photon and many-photon regimes.  
Thus one expects neither a simple perturbative nor the quasistatic 
approximation to be applicable. This ambivalent character is shown to 
be clearly visible, since, e.\,g., the ionization yield shows a 
pronounced dependence on the internuclear separation (as is expected 
for the quasistatic regime), but also clear structures due to 
resonance-enhanced multiphoton ionization (REMPI). However, these 
structures are to a large extent washed out, if one goes beyond the 
fixed-nuclei approximation. In the present work vibrational motion 
is considered in an approximate way where the internuclear-separation 
dependent ionization yields are weighted by the vibrational wavepacket 
of the initial state. The resulting ionization yields of H$_2$ and 
D$_2$ are compared with each other to resolve the isotope effect. 
Since for very short laser pulses the pulse duration influences the 
strong-field behavior, the effects of the laser-pulse duration are  
investigated considering pulse lengths of 5, 10, and 20\,fs. 
In the following atomic units ($e=m_e=\hbar=1$) are used
unless specified otherwise.

%%%%%%%%%%%%%%%%%%%%%%%%%%%%%%%%%%%%%%%%%%%%%%%%%%%%%%%%%%%%%%%%%%%%%%%%%%%%%
%%%%%%%%%%%%%%%%%%%%%%%%%%%%%%%%%%%%%%%%%%%%%%%%%%%%%%%%%%%%%%%%%%%%%%%%%%%%%
\section{\label{sec:Method}Method}
%%%%%%%%%%%%%%%%%%%%%%%%%%%%%%%%%%%%%%%%%%%%%%%%%%%%%%%%%%%%%%%%%%%%%%%%%%%%%
%%%%%%%%%%%%%%%%%%%%%%%%%%%%%%%%%%%%%%%%%%%%%%%%%%%%%%%%%%%%%%%%%%%%%%%%%%%%%

Our method of solving the TDSE describing molecular hydrogen exposed to a
laser field for parallel orientation and its extension to the perpendicular 
one is discussed in detail 
in \cite{sfm:awas05} and \cite{sfm:vann08}, respectively. Briefly, the TDSE
is solved by expanding the time-dependent wave function in terms of
field-free states. The latter are obtained from a 
configuration-interaction (CI) calculation~\cite{dia:vann04} in which 
the Slater determinants are formed with the aid of H$_2^+$ wave functions 
expressed in terms of B splines in prolate spheroidal coordinates 
($1\leq \xi < \infty, -1\leq \eta\leq 1, 0\leq \phi < 2\pi$). 
The use of a B-spline basis confined within 
a finite spatial volume defined by parameter $\xi_{\rm max}$ results in
a suitable discretization of the electronic continuum.

%%%%%%%%%%%%%%%%%%%%%%%%%%%%%%%%%%%%%%%%%%%%%%%%%%%%%%%%%%%%%%%%%%%%%%%%%%%%%
\subsection{\label{subsec:CI}
Configuration-interaction calculation}
%%%%%%%%%%%%%%%%%%%%%%%%%%%%%%%%%%%%%%%%%%%%%%%%%%%%%%%%%%%%%%%%%%%%%%%%%%%%%

%
%%%%%%%%%%%%%%%%%%%%%%%%%%%%%%%%%%%%%%%%%%%%%%%%%%%%%%%%%%%%%%%%%%%%%%%%%%%%%
\begin{figure*}[!tbp]
\begin{center}
\includegraphics[width=0.95\textwidth]{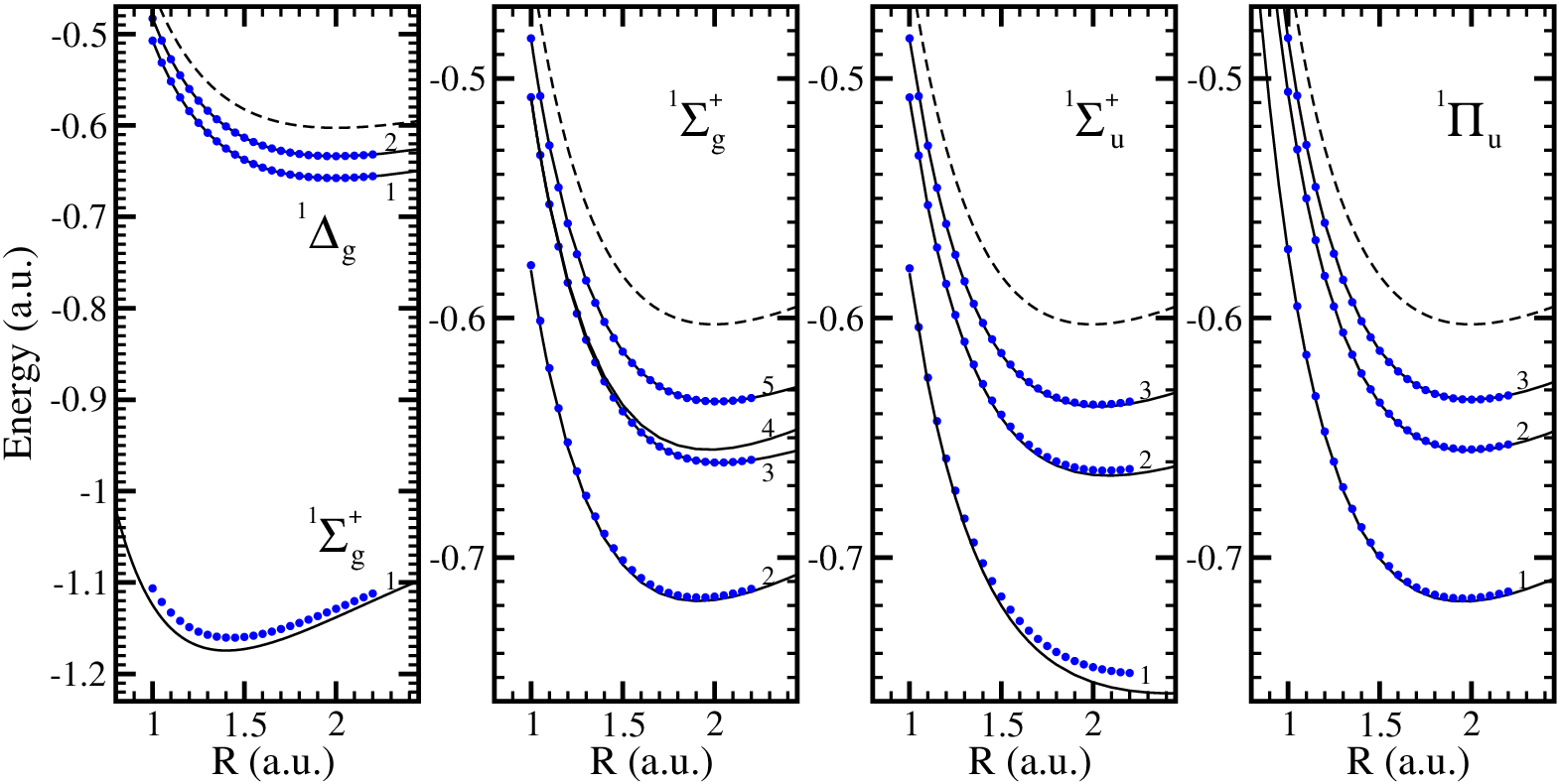}%
\caption{\label{fig:PC_CIvsQE} 
Electronic potential curves of some low-lying states of H$_2$: 
the CI results of this work (blue circles) are compared to the supposedly 
exact values (solid 
lines, ~\cite{dia:woln93,dia:woln94,dia:dres95,dia:stas02,dia:woln03b}).  
The dashed line shows the ionization threshold of H$_2$. 
The different adiabatic electronic states of a given molecular symmetry 
are numbered in the order of the energy.} 
\end{center}
\end{figure*}
%%%%%%%%%%%%%%%%%%%%%%%%%%%%%%%%%%%%%%%%%%%%%%%%%%%%%%%%%%%%%%%%%%%%%%%%%%%%%
%

For the considered laser parameters the electronic field-free states
for every molecular symmetry were obtained as follows. 
A box size  of about 350\,$a_0$ is chosen 
independently of $R$. This is achieved by a proper adaption of 
$\xi_{\rm max}$ as a function of $R$.  
Along the $\xi$ coordinate 350 $B$ splines of 
order $k=10$ with an almost linear knot sequence were used where the 
first 40 intervals are increased by a geometric progression using a 
progression factor $g=1.05$ and then the size of the interval remains 
constant. Along the $\eta$ coordinate 30 $B$ splines of order 8 were 
used in the complete 
interval $-1\,\leq \eta \leq 1$, but using the symmetry of a 
homonuclear system as is described in~\cite{dia:vann04}. 
Out of the resulting 5235 orbitals for every symmetry only 3490 orbitals
were further used to construct CI configurations, whereas those orbitals 
with highly oscillating angular part (with more than 19 nodes for the 
$\eta$-dependent component) were omitted. In most of the subsequent CI 
calculations approximately 6000 configurations were used for every symmetry. 
These states result from a very long configuration series (3490 
configurations) in which one electron occupies the H$_2^+$ ground-state 
$1\,\sigma_g$ orbital while the other one is occupying one of the 
remaining, e.\,g., $n\,\pi_u$ or $n\,\delta_g$ orbitals. The other 
CI configurations represent doubly excited situations and are 
responsible for describing correlation (and real doubly excited states).
Finally, out of the obtained CI states only those with an energy  below
the energy cut-off (chosen at 10\,a.u.\ above the ionization threshold for
the calculations shown in this work) were included in the time propagation 
(about 5400 states per symmetry).
For the perpendicular orientation only molecular symmetries with the
absolute value of the component of the total angular momentum along the
internuclear axis $0\leq \Lambda \leq 7$ were included in the time 
propagation. This results in a system of about 86,000 real-valued 
first-order differential equations. Noteworthy, the adopted range of 
$\Lambda$ values does not only guarantee  
the convergence of ionization yields, but also provides a reasonable 
description of photoelectron energy spectra.

The basis set specified above was chosen to provide a good compromise for 
describing a large
number of states and can, of course, not compete with a high-precision 
calculation optimized for a single electronic state. Fig.~\ref{fig:PC_CIvsQE}
demonstrates a comparison of the obtained electronic energies for different 
low-lying molecular states with high-precision calculations of
L. Wolniewicz and co-workers performed using an explicitly correlated 
basis~\cite{dia:woln93,dia:woln94,dia:dres95,dia:stas02,dia:woln03b}.
(The present CI method is able to reproduce such practically exact 
electronic energies at least within 4-6 significant digits, if the basis 
set is chosen judiciously~\cite{dia:vann04}). The agreement is excellent 
for all states except for 1${}^1\Sigma^{+}_g$ 
and 1${}^1\Sigma^{+}_u$ (at larger internuclear distances) where
the electronic motion is highly correlated and cannot efficiently be 
described by a CI calculation employing orbitals with no electron-electron 
interaction included. Nevertheless, even for these two states
the obtained electronic energies are much better than those obtained with 
the Hartree-Fock approximation. For example, for the ground state of H$_2$ 
with the exact electronic energy at $R=1.4\,a_0$ being equal to
$-1.1745\,$a.u., the Hartree-Fock limit is $-1.1336\,$a.u., whereas the 
present CI calculation yields $-1.1604\,$a.u.

%%%%%%%%%%%%%%%%%%%%%%%%%%%%%%%%%%%%%%%%%%%%%%%%%%%%%%%%%%%%%%%%%%%%%%%%%%%%%
\subsection{\label{subsec:Integr}
Integration over internuclear separations}
%%%%%%%%%%%%%%%%%%%%%%%%%%%%%%%%%%%%%%%%%%%%%%%%%%%%%%%%%%%%%%%%%%%%%%%%%%%%%

Once the TDSE is solved for a given linear-polarized laser pulse, fixed 
internuclear separation $R$, and angle $\theta$ between the internuclear 
and the polarization axis, the ionization yield $Y_{\rm ion}(R,\theta)$ 
is obtained from a summation over the populations of all discretized 
continuum states. These yields can be further used to calculate 
the ionization yield $Y_{\rm ion}^{(\nu)}(\theta)$
for a given initial vibrational state $\nu$ described by the vibrational 
wavefunction $\phi_{\nu}(R)$. Indeed, if the duration of the pulse is 
sufficiently short and depletion of the state during the pulse is 
insignificant, one can neglect the motion of the wavepacket created 
during the pulse and calculate the total ionization yield 
as~\cite{sfm:saen00c}%
%---------------
\begin{equation}
 Y_{\rm ion}^{(\nu)}(\theta) = 
            \int \md R \,Y_{\rm ion}(R,\theta)\, |\phi_{\nu}(R)|^2 \quad .
\label{eq:Yionvibr}
\end{equation}
%---------------
Application of Eq.~(\ref{eq:Yionvibr}) is further based on the assumption 
that the molecule has no time to rotate during the 
pulse and neglect distortion of the electronic ground-state potential 
curve due to the external field. The latter assumption implies that the 
index $\nu$ of vibrational state should be sufficiently small. In this 
work, only $\nu=0$ is considered, but for the two isotopes H$_2$ and 
D$_2$ with their different vibrational wavefunctions. Due to the 
larger mass the D$_2$ vibrational ground state is more compact 
than the one of H$_2$.

%%%%%%%%%%%%%%%%%%%%%%%%%%%%%%%%%%%%%%%%%%%%%%%%%%%%%%%%%%%%%%%%%%%%%%%%%%%%%
\subsection{\label{subsec:AModel}Atomic model}
%%%%%%%%%%%%%%%%%%%%%%%%%%%%%%%%%%%%%%%%%%%%%%%%%%%%%%%%%%%%%%%%%%%%%%%%%%%%%

\begin{figure}
\begin{center}
\includegraphics[width=8cm]{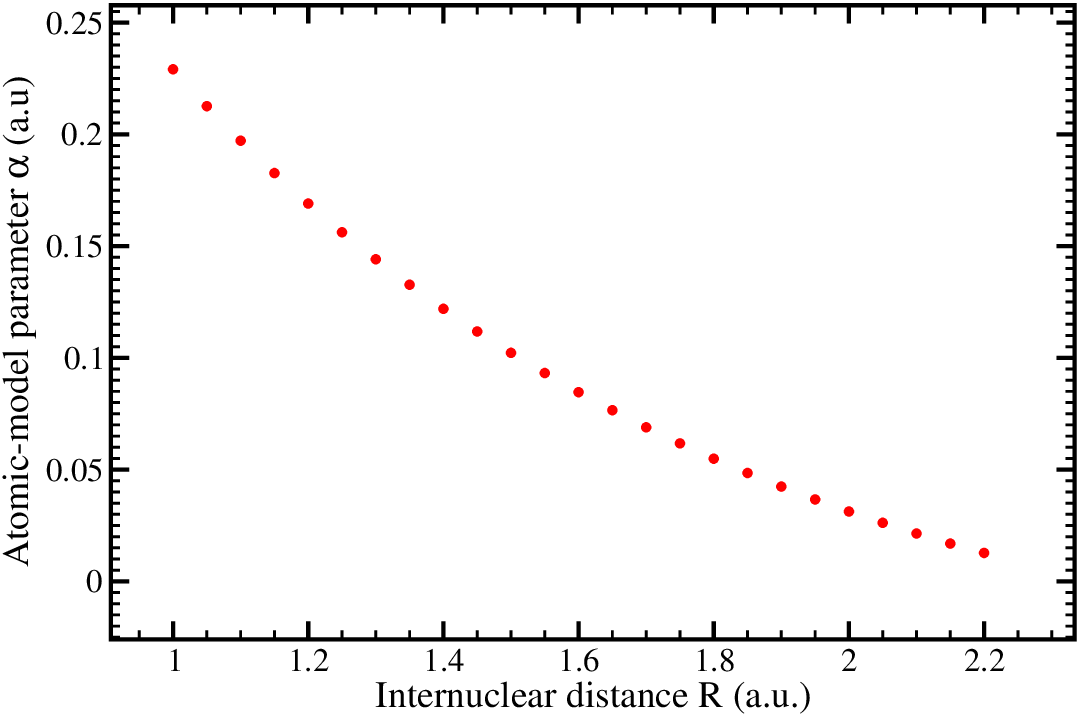}%
\caption{\label{fig:H2_400nm_fig_2} 
Dependence of the atomic-model parameter $\alpha$ [see Eq.~(\ref{eq:Vatom})] 
on the internuclear distance $R$. This choice of $\alpha(R)$ leads to a 
vertical ionization potential of H$_2$ in agreement with the one of the 
present CI calculation.}
\end{center}
\end{figure}

For the
analysis of the orientational dependence of the ionization due to 
the anisotropy of a molecule, it is convenient to compare the molecular 
results with those obtained for an artificial atom with an isotropic,
single-centered charge distribution. 
Strong-field ionization is, however, known to be not only sensitive 
to the symmetry, but also to the electronic binding energy  
and the exact form of the long-ranged Coulomb potential. Therefore, 
the artificial atom must agree to the corresponding molecule 
with respect to the two latter factors. For this reason the simple 
single-electron one-parameter model potential
\begin{equation}
\label{eq:Vatom}
  V(r) = -\frac{1}{r} \left\{ 1 + \frac{\alpha}{|\alpha|} 
          \exp\left[- \frac{2r}{|\alpha|^{1/2}}\right] \right\}
\end{equation}
was introduced~\cite{sfm:vann08}. Its performance for describing 
various physical problems was checked in \cite{dia:luhr08} and it was 
recently also applied to the calculation of antiproton--H$_2$ scattering 
cross sections and stopping powers \cite{anti:luhr08a,anti:luhr09c}. The 
ionization potential $I_p$ of such an artificial atom is directly 
related to the parameter
$\alpha$ of the model potential (\ref{eq:Vatom}). For a given ionization
potential $I_p$ the corresponding parameter $\alpha$ can be found as 
$\alpha = \alpha(I_p)$, where $\alpha(I_p)$ can be obtained numerically.
Since the molecular vertical ionization potential (energy difference 
between the electronic ground-state potential curves of the ion and the 
neutral) depends on the internuclear
distance $R$, the value of $\alpha$ should also depend on $R$
in order to compare atomic-model and molecular results.
Figure \ref{fig:H2_400nm_fig_2} shows the $R$ dependence of $\alpha$. 
It is determined by requiring the resulting $R$-dependent ionization 
potential to agree with the one obtained by the present CI calculation 
for the H$_2$ ground state. 

%
%%%%%%%%%%%%%%%%%%%%%%%%%%%%%%%%%%%%%%%%%%%%%%%%%%%%%%%%%%%%%%%%%%%%%%%%%%%%%
%%%%%%%%%%%%%%%%%%%%%%%%%%%%%%%%%%%%%%%%%%%%%%%%%%%%%%%%%%%%%%%%%%%%%%%%%%%%%
%

In order to compare to the full molecular calculations the atomic-model 
results are multiplied  by a factor 2 that accounts for the two equivalent
electrons in molecular hydrogen. This procedure is known to be reasonable for
ionization yields less than 10-20\%~\cite{sfm:vann08} which is the case for
the present calculations.

%%%%%%%%%%%%%%%%%%%%%%%%%%%%%%%%%%%%%%%%%%%%%%%%%%%%%%%%%%%%%%%%%%%%%%%%%%%%%
%%%%%%%%%%%%%%%%%%%%%%%%%%%%%%%%%%%%%%%%%%%%%%%%%%%%%%%%%%%%%%%%%%%%%%%%%%%%%
\section{\label{sec:Results}Results}
%%%%%%%%%%%%%%%%%%%%%%%%%%%%%%%%%%%%%%%%%%%%%%%%%%%%%%%%%%%%%%%%%%%%%%%%%%%%%
%%%%%%%%%%%%%%%%%%%%%%%%%%%%%%%%%%%%%%%%%%%%%%%%%%%%%%%%%%%%%%%%%%%%%%%%%%%%%

All calculations presented in this work were performed with $N$-cycle 
cos$^2$-shaped linear-polarized laser pulses with $N = 10, 20$, and 40.
For a wavelength of 400\,nm the FWHM of intensity of such pulses 
corresponds to about 5, 10, and 20\,fs.

%%%%%%%%%%%%%%%%%%%%%%%%%%%%%%%%%%%%%%%%%%%%%%%%%%%%%%%%%%%%%%%%%%%%%%%%%%%%%
\subsection{\label{subsec:FIR}Field-induced resonances}
%%%%%%%%%%%%%%%%%%%%%%%%%%%%%%%%%%%%%%%%%%%%%%%%%%%%%%%%%%%%%%%%%%%%%%%%%%%%%

%
%%%%%%%%%%%%%%%%%%%%%%%%%%%%%%%%%%%%%%%%%%%%%%%%%%%%%%%%%%%%%%%%%%%%%%%%%%%%%
\begin{figure}
\begin{center}
\includegraphics[width=8cm]{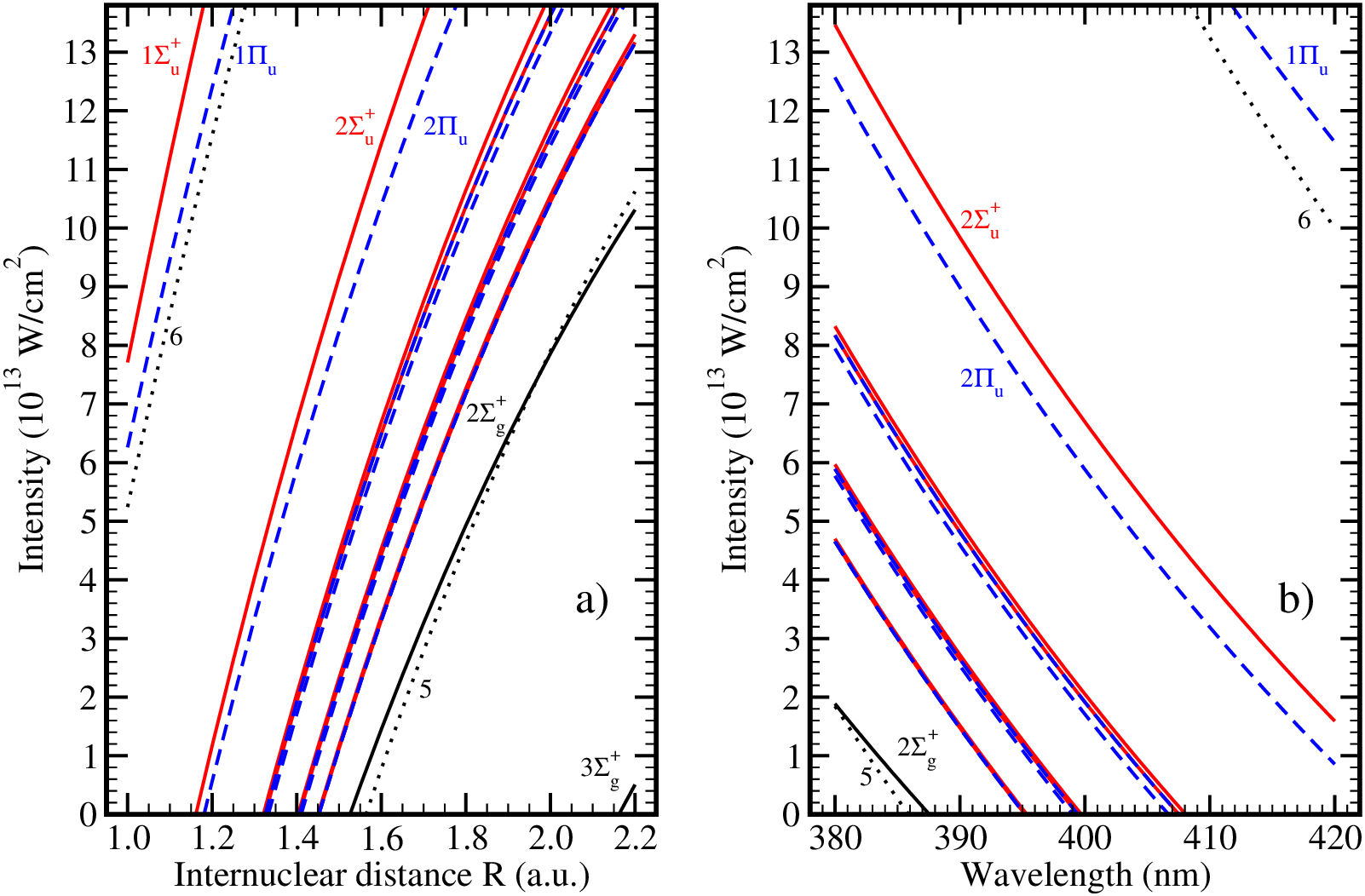}%
\caption{\label{fig:H2_400nm_fig_3} 
Expected positions of REMPI peaks and $N$-photon ionization thresholds.
Dependence of the resonant intensities $I$ (a) on the internuclear 
distance $R$ for pulses with a wavelength of 400\,nm and 
(b) on wavelength for the fixed internuclear distance $R=1.4\,a_0$. 
Positions of REMPI peaks due to $\Sigma_u^+$ (red solid), 
$\Pi_u$ (blue dashes), or $\Sigma_g^+$ (black solid) 
intermediate states are shown. Also shown are the $N$-photon ionization  
thresholds (dotted lines, with $N$ explicitly indicated in the graph).} 
\end{center}
\end{figure}
%%%%%%%%%%%%%%%%%%%%%%%%%%%%%%%%%%%%%%%%%%%%%%%%%%%%%%%%%%%%%%%%%%%%%%%%%%%%%
%

For the following discussion of the orientation dependence and isotope 
effects, it is helpful to obtain a more
detailed understanding of the influence of such parameters as peak intensity
and wavelength of the pulse or of the internuclear separation on the
positions of REMPI peaks and $N$-photon ionization thresholds. In order to
be able to correctly predict REMPI through some resonant electronic state 
in intense laser pulses, it is necessary to know the field-induced shift of the
resonant state, what is a challenging task by itself. However, if the field is
sufficiently intense, one can assume that the field-induced shift (dynamically
induced Stark shift) of excited 
states is almost equal to the ponderomotive energy. With this assumption 
and using the field-free transition energies of the present CI calculation  
the positions of the REMPI peaks are expected to depend on the laser 
parameters and internuclear separation as shown in 
Fig.~\ref{fig:H2_400nm_fig_3}. 

The dependence of the expected positions of the REMPI peaks and
$N$-photon ionization thresholds on the 
internuclear distance $R$ for a 400\,nm laser field is given  
in Fig.~\ref{fig:H2_400nm_fig_3}\,a. In this $R$ range 
the ionization process can be referred to as
5-photon (7-photon) ionization in the bottom-right (top-left) part of 
the figure, or as 6-photon ionization otherwise. Different kinds of 
REMPI peaks are expected: (5+1) REMPI peaks through 
$n \Sigma_u^+$ or $n \Pi_u$ electronic states with $n>1$, 
(5+2) REMPI peaks through the $1 \Sigma_u^+$ or $1 \Pi_u$  
states, a (4+1) REMPI peak through the $3\Sigma_g^+$ state, 
and a REMPI peak through the $2\Sigma_g^+$ state. 
Note, that in the last case the expected position of the resonance 
crosses the expected position of the 5-photon ionization threshold. 
Therefore, the resonance can be
referred to as (4+1) REMPI for peak intensities smaller than 
$7.5\times 10^{13}$\,W/cm$^2$, and as (4+2) REMPI for higher intensities. 
Evidently, the correct character of the resonance is sensitive to the exact
intensity dependence of the field-induced shift of the $2\Sigma_g^+$
state, and thus a non-trivial behavior is expected. A similar
conclusion is valid for the REMPI through the $1 \Sigma_u^+$ or 
$1 \Pi_u$ electronic states, since their exact REMPI positions could  
in fact cross the 6-photon ionization threshold. 

\begin{figure}[!tbp]
\centering
\subfigure[\quad$R=1.4\,a_0$ ]{
\includegraphics[width=4.1cm]{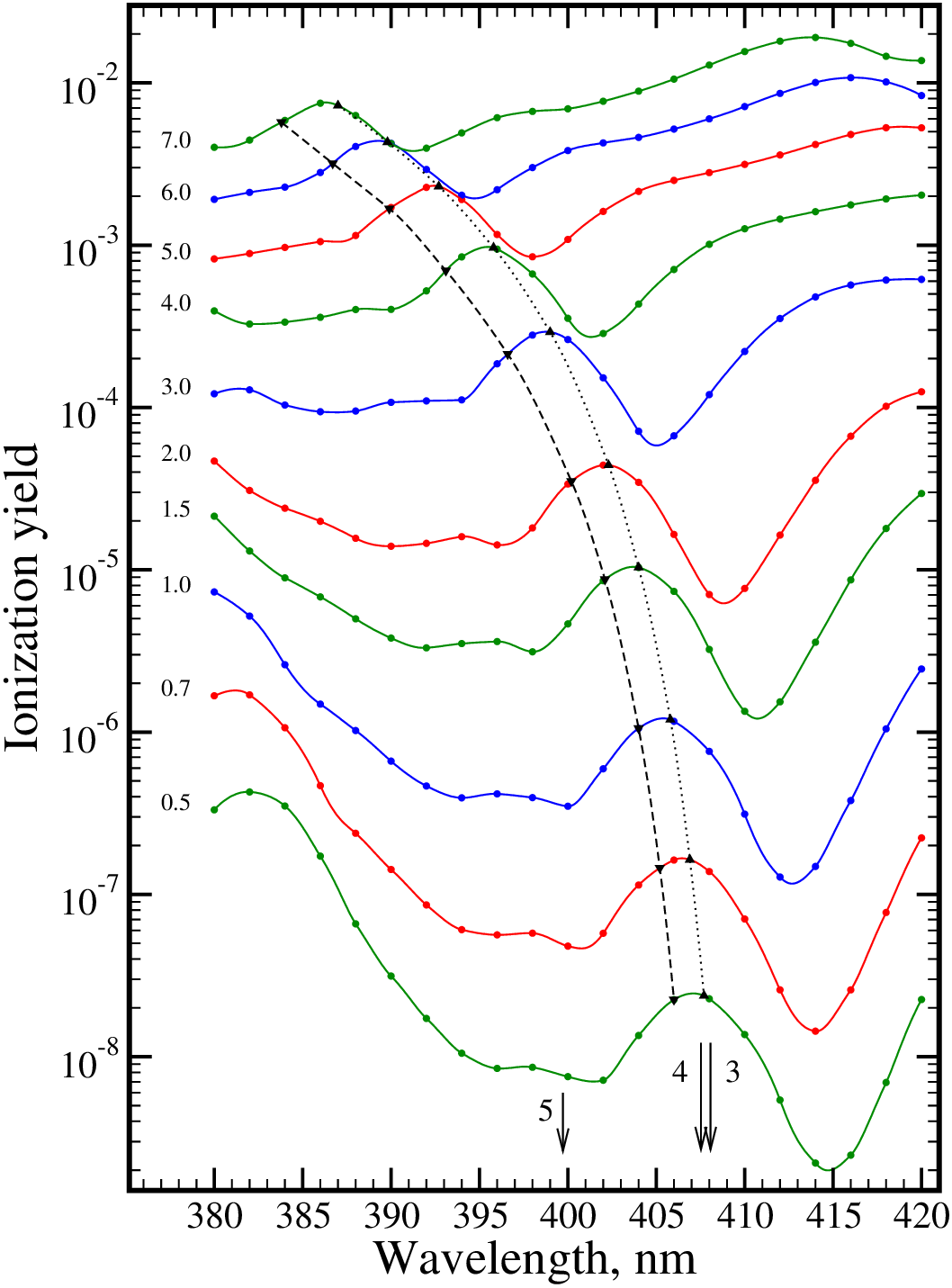}%
}
\subfigure[\quad$R=2.2\,a_0$ ]{
\includegraphics[width=4.1cm]{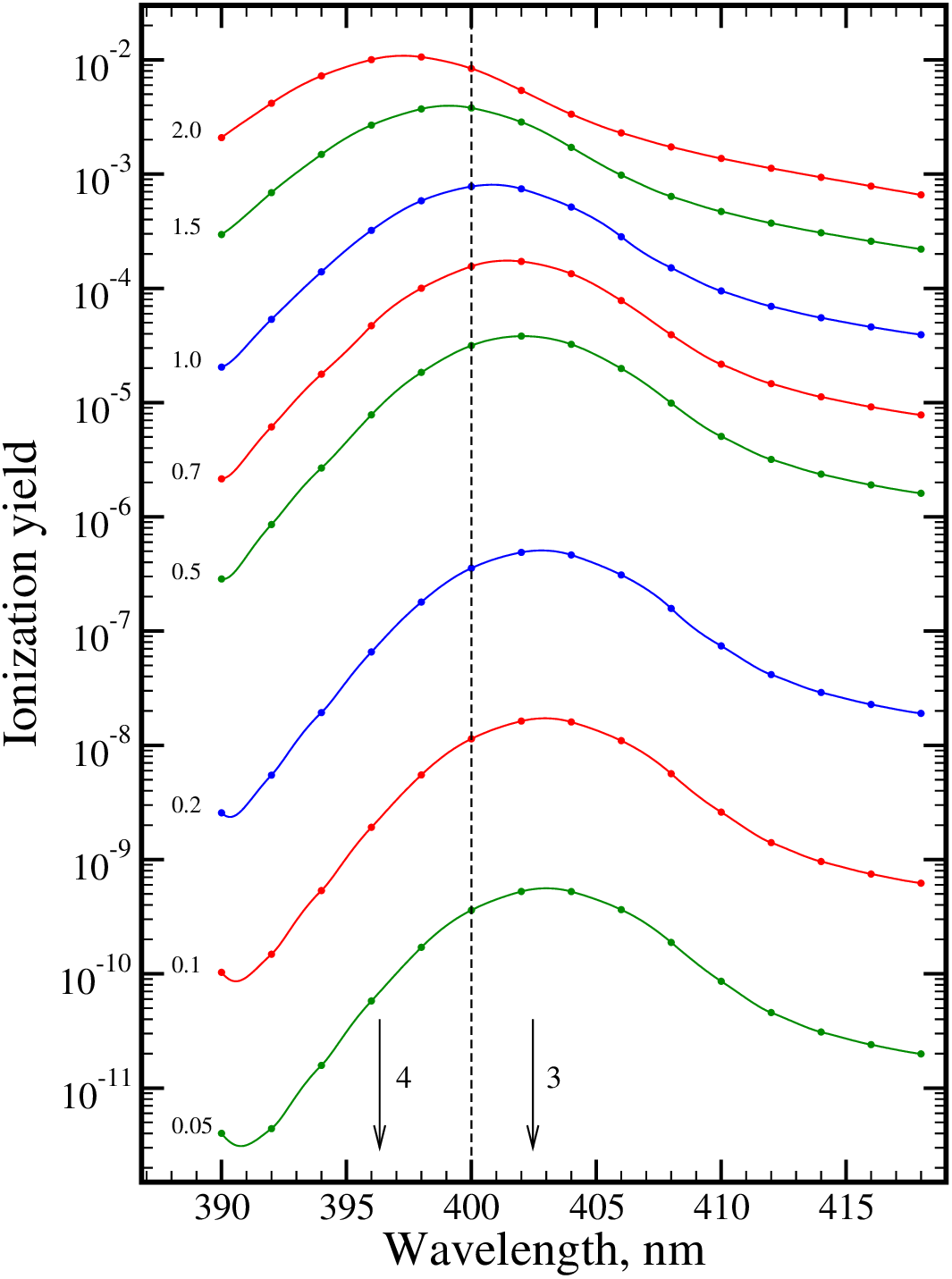}%
}
\caption{\label{fig:Ion_H2R-aS_C-gxxa} 
Wavelength-dependent ionization yields for a parallel orientation of 
an H$_2$ molecule at the fixed internuclear distances 
(a) $R=1.40\,a_0$ and (b) $R=2.20\,a_0$ 
for various peak intensities 
(specified in units of $10^{13}\,$W/cm$^2$) of 
40-cycle cos$^2$-shaped pulses. The arrows indicate the resonant 
wavelengths (in the low-intensity limit) of REMPI peaks due to  
$n \Sigma_u^+$ (a) and $n \Sigma_g^+$ (b) intermediate states. 
The expected (dashes) and found (dots) position of a REMPI 
peak is also given in (a).} 
\end{figure}

Similarly, Fig.~\ref{fig:H2_400nm_fig_3}\,b shows the dependence of
expected positions of REMPI peaks and $N$-photon ionization thresholds
on the laser wavelength for the fixed internuclear
distance $R=1.4\,a_0$. 
With larger peak intensity the increasing ponderomotive energy leads
to an increase of the transition energy between the initial and the 
resonant state. This increase can be compensated by the increase of the 
photon energy, and thus the new REMPI position will occur at a smaller 
wavelength. The calculations at a fixed internuclear separation are 
more suitable for the investigation of the validity of the assumed 
field-induced shift of electronic states, since in this case identical  
sets of field-free electronic wavefunctions are used in the time 
propagation. For this
purpose, a series of 210 calculations for a parallel-oriented H$_2$ 
molecule with fixed internuclear distance $R=1.4\,a_0$ exposed to 
laser pulses with a total duration of 40 cycles was performed for 
21 different values of the wavelength and 10 different values of 
the peak intensity. The results are shown in 
Fig.~\ref{fig:Ion_H2R-aS_C-gxxa}\,a, where every point represents 
the outcome of one full TDSE calculation, and curves join the results 
obtained for the same peak intensity.

%
%%%%%%%%%%%%%%%%%%%%%%%%%%%%%%%%%%%%%%%%%%%%%%%%%%%%%%%%%%%%%%%%%%%%%%%%%%%%%%%%%%%%%
%%%%%%%%%%%%%%%%%%%%%%%%%%%%%%%%%%%%%%%%%%%%%%%%%%%%%%%%%%%%%%%%%%%%%%%%%%%%%%%%%%%%%
%

Figure \ref{fig:Ion_H2R-aS_C-gxxa}\,a shows a pronounced peak 
whose position moves from 407\,nm for a peak intensity
of $5 \times 10^{12}$\,W/cm$^2$ to 387\,nm for a peak intensity of
$7 \times 10^{13}$\,W/cm$^2$. From Fig.~\ref{fig:H2_400nm_fig_3}\,b
it follows that this peak can be assigned to REMPI through either one or 
both of the closely lying $3
\Sigma_u^+$ and $4 \Sigma_u^+$ electronic states, since the
spectral width of the Fourier-limited pulse is too broad to resolve 
these two resonances. Clearly, the intensity-dependent shift of the 
peak position is overestimated by the already mentioned simple 
prediction based purely on the ponderomotive energy ($\delta E(I) = U_p$). 
Instead, the found intensity dependence of the field-induced energy shift 
can be well fitted by $\delta E(I) = 0.9 U_p - 0.002$.  
At intensities $5 \times 10^{12}$\,W/cm$^2$ and smaller it appears 
as the energy shift of these low-lying excited states (responsible 
for the REMPI) is already absent. The position of the REMPI peak 
agrees then much better with the low-intensity limit than with 
the prediction based on $U_p$, since the latter would predict 
a shift of about 2\,nm. Such a shift by about 2\,nm is, however, 
found for the (poorly resolved) REMPI peak due to the 
higher lying $5 \Sigma_u^+$ intermediate state. This demonstrates 
that in the investigated regime of laser parameters different 
excited states behave differently, and a common prediction for 
all excited states is impossible.  

Figure \ref{fig:Ion_H2R-aS_C-gxxa}\,b shows again the results of a 
series of (this time 120) calculations for a parallel-oriented H$_2$ 
molecule, but for the larger internuclear separation $R=2.2\,a_0$. 
In this case the spacing of the $3 \Sigma_g^+$ and $4 \Sigma_g^+$ 
states that could lead to (4+1) REMPI is rather large, and thus 
the pronounced peak in Fig.~\ref{fig:Ion_H2R-aS_C-gxxa}\,b
can be entirely assigned to REMPI through the $3 \Sigma_g^+$ state. 
Although the peak position clearly shifts to smaller wavelengths 
with increasing laser peak intensity, the
shift becomes visible only for rather large intensities. As a consequence,
the peak position crosses 400\,nm at an intensity higher than 
$10^{13}$\,W/cm$^2$, whereas according to
Fig.~\ref{fig:H2_400nm_fig_3}\,a the crossing should have occurred at
an intensity that is smaller by a factor 2. Thus, although the 
positions presented in
Fig.~\ref{fig:H2_400nm_fig_3} give a satisfactory explanation of the
main features, they should only be considered as a rough estimate.

%%%%%%%%%%%%%%%%%%%%%%%%%%%%%%%%%%%%%%%%%%%%%%%%%%%%%%%%%%%%%%%%%%%%%%%%%%%%%
\subsection{$R$-dependent ionization}
%%%%%%%%%%%%%%%%%%%%%%%%%%%%%%%%%%%%%%%%%%%%%%%%%%%%%%%%%%%%%%%%%%%%%%%%%%%%%

%
%%%%%%%%%%%%%%%%%%%%%%%%%%%%%%%%%%%%%%%%%%%%%%%%%%%%%%%%%%%%%%%%%%%%%%%%%%%%%
\begin{figure*}[!tbp]
\centering
\subfigure[Parallel orientation of the internuclear axis to a linear
 polarized laser field]{
\includegraphics[width=16.0cm]{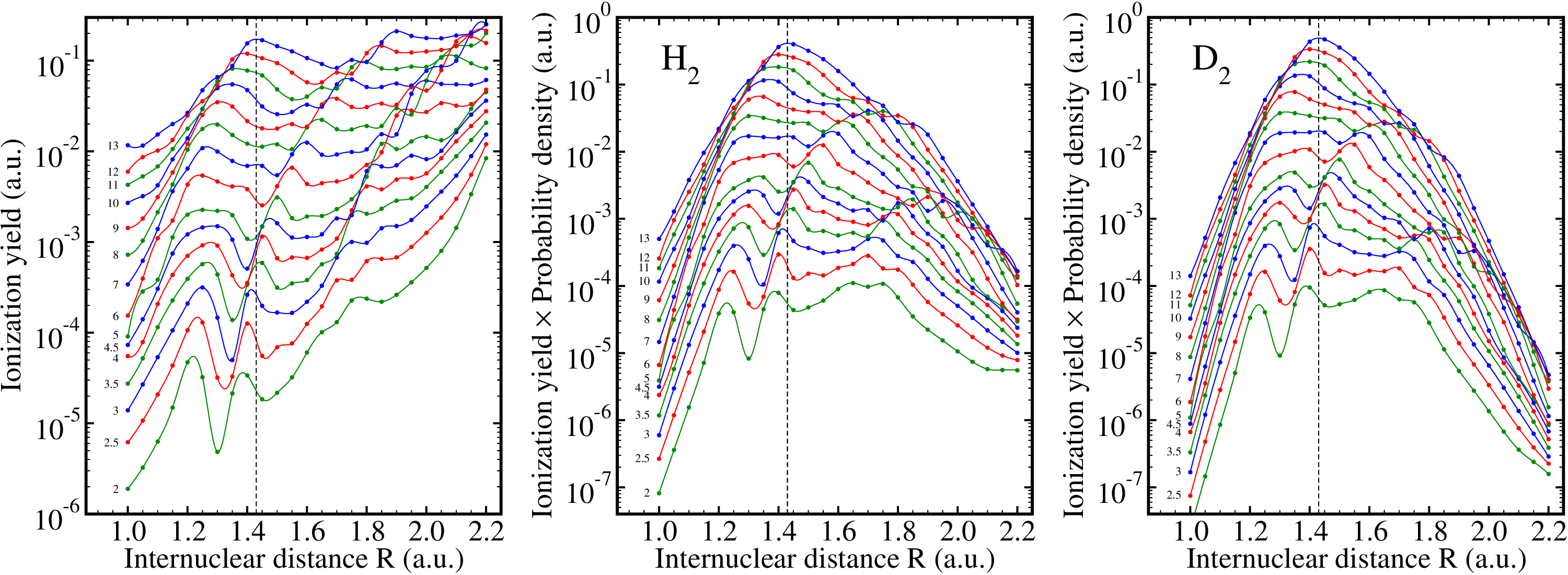}%
}\\
\subfigure[Perpendicular orientation of the internuclear axis to a linear 
polarized laser field]{
\includegraphics[width=16.0cm]{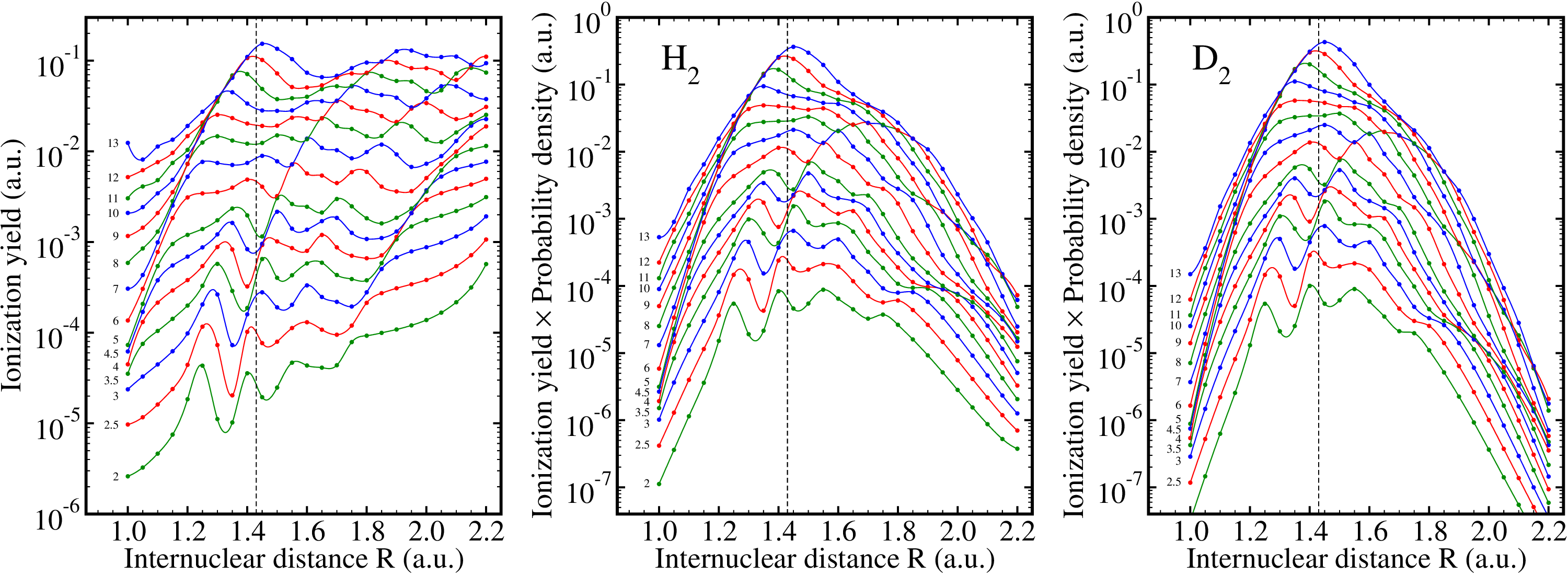}%
}\\
\subfigure[Atomic model calculations (multiplied by a factor 2)]{
\includegraphics[width=16.0cm]{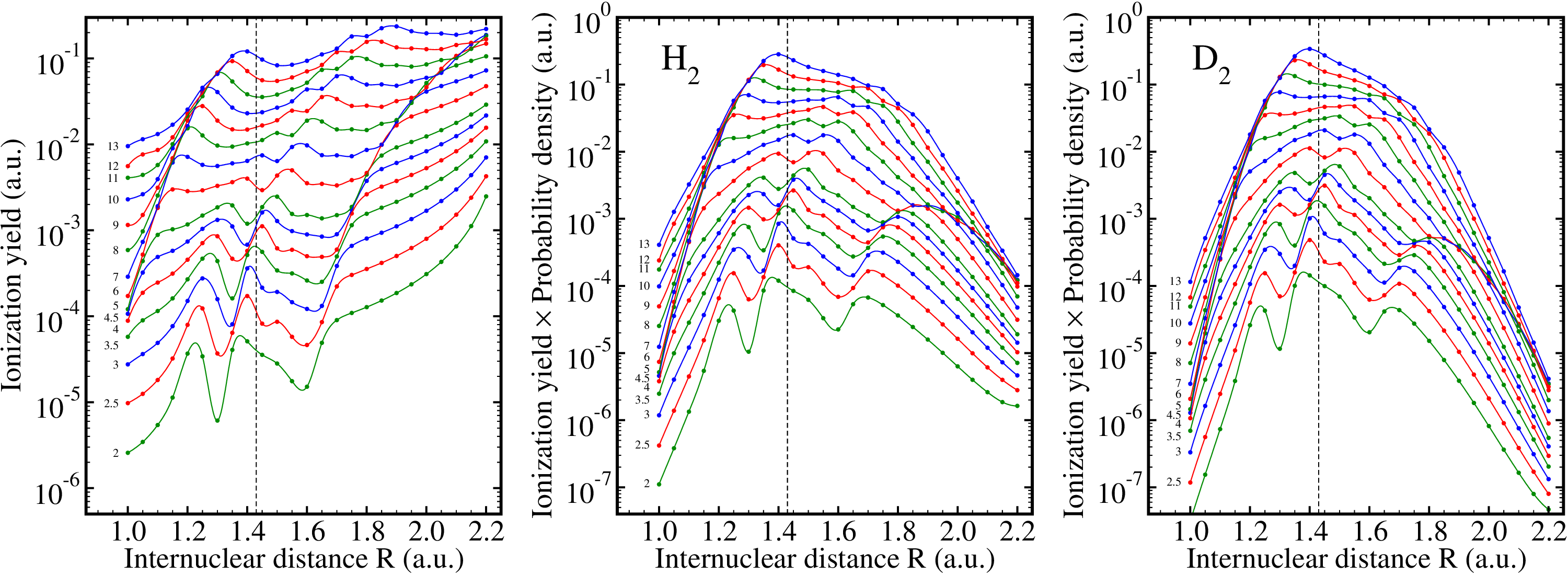}%
}%
\caption{\label{fig:Ion_H2D2_All_CCgxxa} 
Ionization yields for 40-cycle cos$^2$-shaped (20\,fs) laser pulses with 
a wavelength of 400\,nm and different peak intensities in between 
2.0 and 13$\,\times10^{13}\,$W/cm$^2$ (specified in the
graphs in units of $10^{13}\,$W/cm$^2$). The upper (middle) panel shows 
the results for a parallel (perpendicular) orientation of the molecule with 
respect to the field axis, while the lower panel shows the results obtained 
with the atomic model potential in Eq.~(\ref{eq:Vatom}).  
The left panel shows the fixed-nuclei ionization yields as a function of the
internuclear distance $R$, whereas the middle and right panels display the 
ionization yields multiplied with the probability 
density of the ground vibrational states of H$_2$ and D$_2$, respectively. 
(Every point corresponds to a full solution of the TDSE, while the points 
are connected by spline interpolating curves.)} 
\end{figure*}
%%%%%%%%%%%%%%%%%%%%%%%%%%%%%%%%%%%%%%%%%%%%%%%%%%%%%%%%%%%%%%%%%%%%%%%%%%%%%
%
%
%%%%%%%%%%%%%%%%%%%%%%%%%%%%%%%%%%%%%%%%%%%%%%%%%%%%%%%%%%%%%%%%%%%%%%%%%%%%%
\begin{figure*}[!tbp]
\centering
\subfigure[Parallel orientation of the internuclear axis to a linear 
polarized laser field]{
\includegraphics[width=16.0cm]{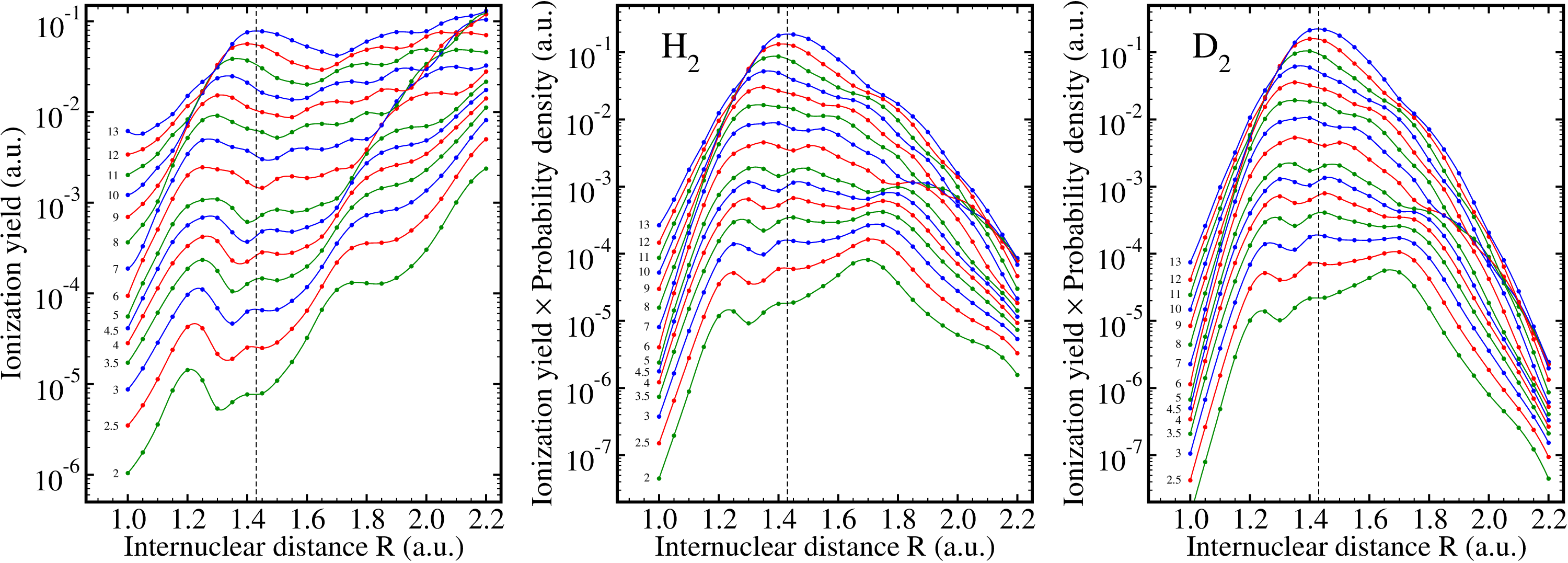}%
}\\
\subfigure[Perpendicular orientation of the internuclear axis to a linear 
polarized laser field]{
\includegraphics[width=16.0cm]{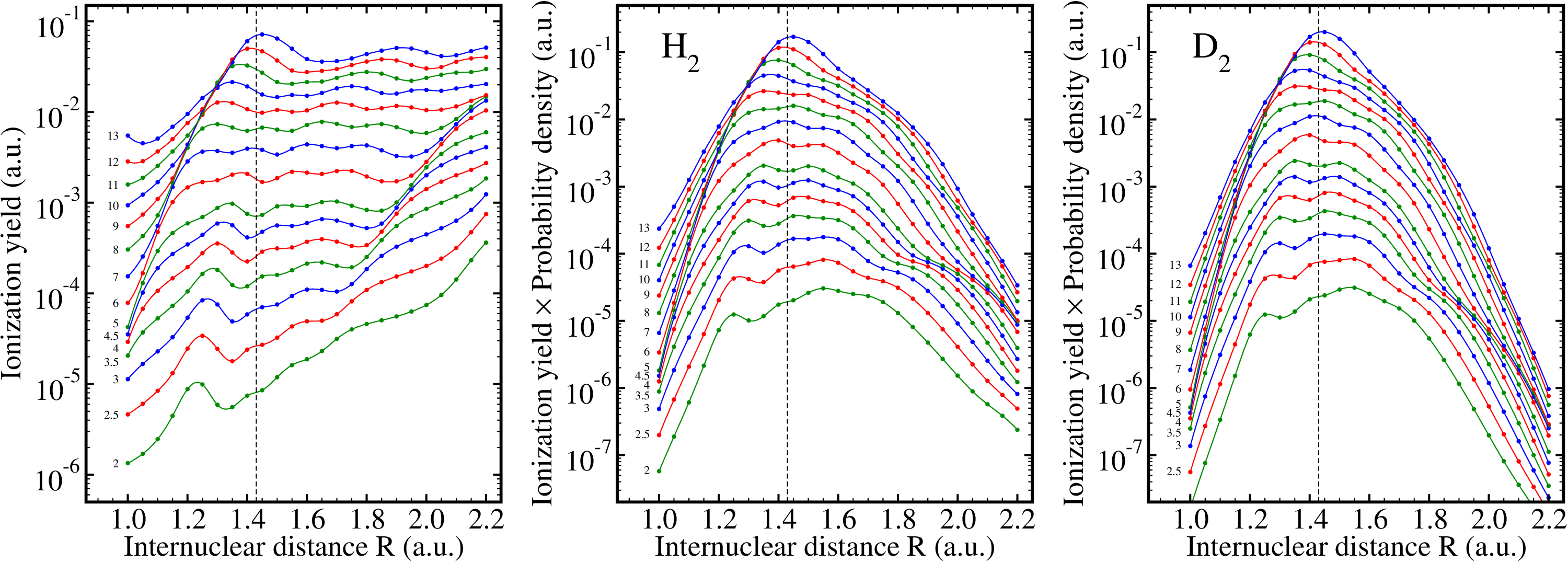}%
}\\
\subfigure[Atomic model calculations (multiplied by a factor 2)]{
\includegraphics[width=16.0cm]{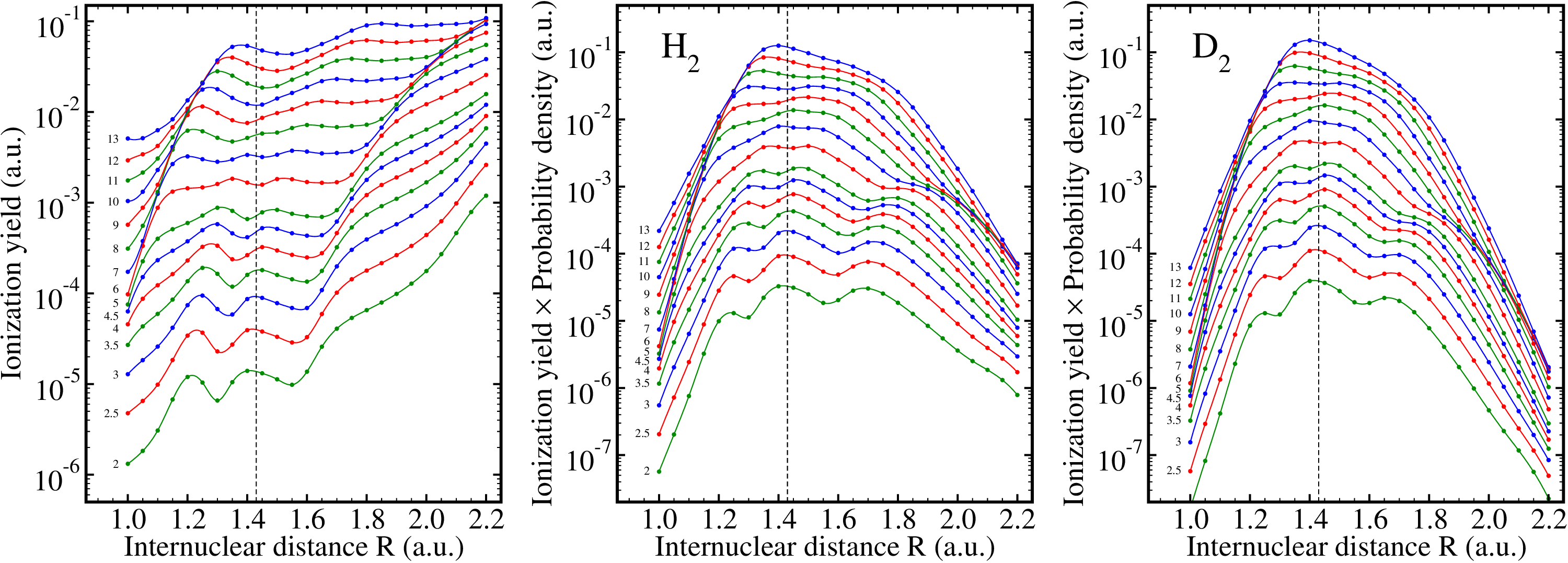}%
}%
\caption{\label{fig:Ion_H2D2_All_CCcxxa} 
As Figure~\ref{fig:Ion_H2D2_All_CCgxxa}, but for a 20-cycle cos$^2$-shaped
(10\,fs) laser pulse.} 
\end{figure*}
%%%%%%%%%%%%%%%%%%%%%%%%%%%%%%%%%%%%%%%%%%%%%%%%%%%%%%%%%%%%%%%%%%%%%%%%%%%%%
%
%
%%%%%%%%%%%%%%%%%%%%%%%%%%%%%%%%%%%%%%%%%%%%%%%%%%%%%%%%%%%%%%%%%%%%%%%%%%%%
\begin{figure*}[!tbp]
\centering
\subfigure[Parallel orientation of the internuclear axis to a linear 
polarized laser field]{
\includegraphics[width=16.0cm]{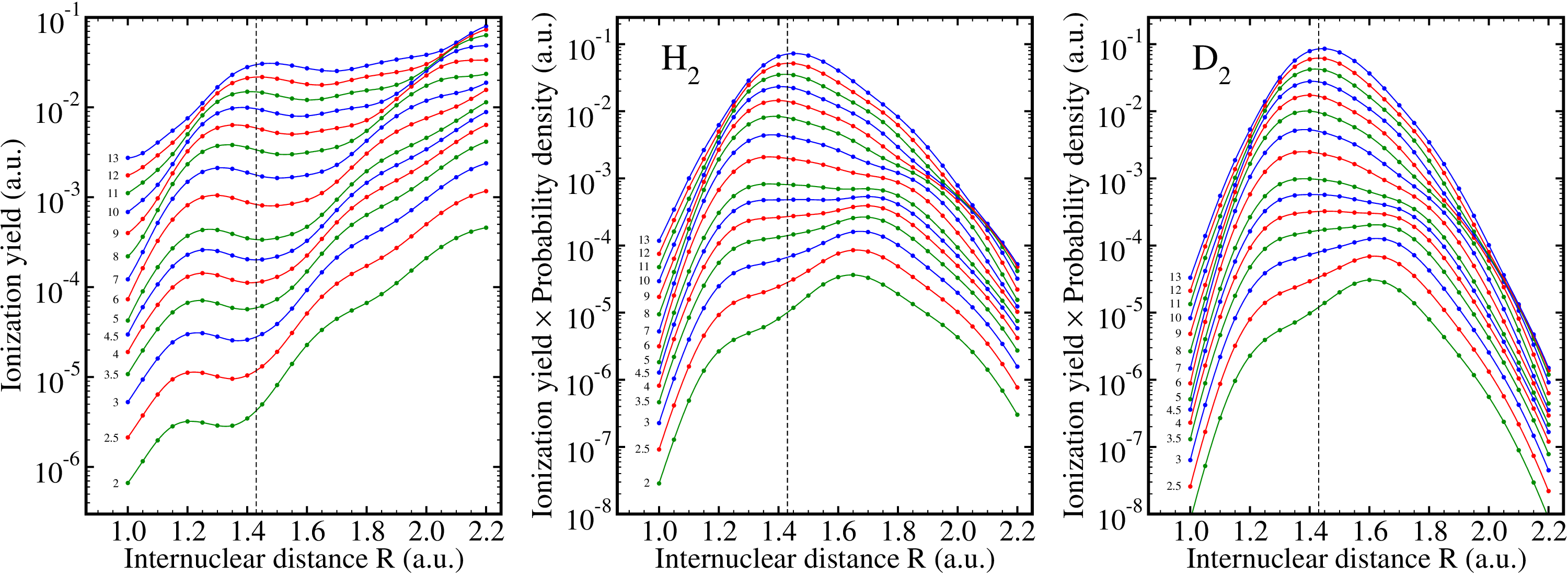}%
}\\
\subfigure[Perpendicular orientation of the internuclear axis to a linear 
polarized laser field]{
\includegraphics[width=16.0cm]{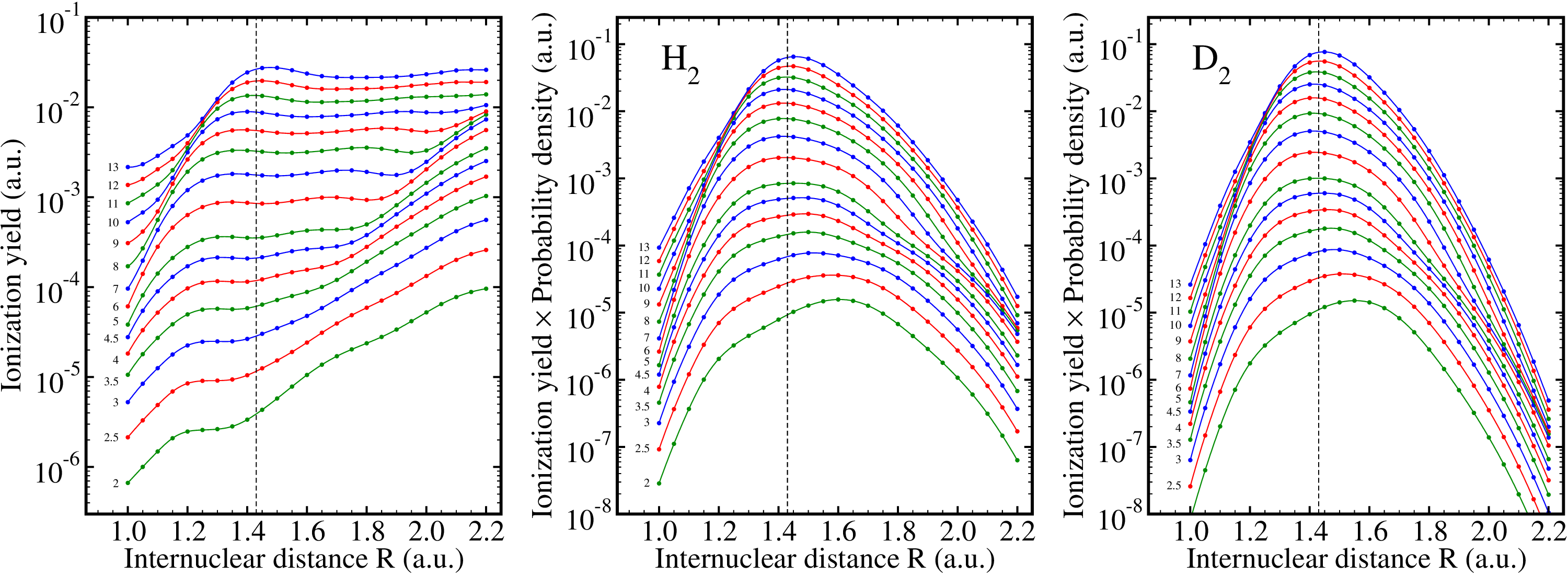}%
}\\
\subfigure[Atomic model calculations (multiplied by a factor 2)]{
\includegraphics[width=16.0cm]{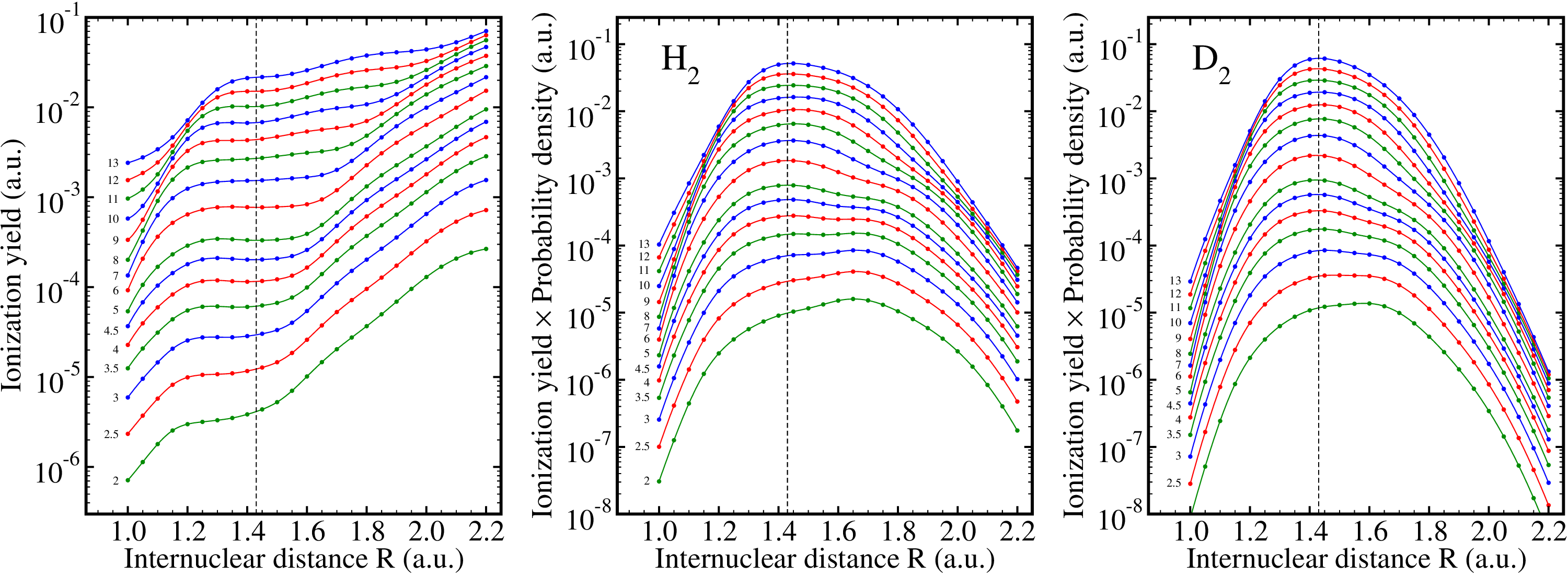}%
}%
\caption{\label{fig:Ion_H2D2_All_CC5xxa} 
As Figure~\ref{fig:Ion_H2D2_All_CCgxxa}, but for a 10-cycle cos$^2$-shaped
(5\,fs) laser pulse.} 
\end{figure*}
%%%%%%%%%%%%%%%%%%%%%%%%%%%%%%%%%%%%%%%%%%%%%%%%%%%%%%%%%%%%%%%%%%%%%%%%%%%%%
%

If strong-field ionization of H$_2$ or D$_2$ initially in their vibrational 
ground states is considered, it is important to investigate the dependence 
of the ionization on the internuclear separation $R$ within an $R$ range 
in which the vibrational wave function is nonvanishing (Franck-Condon 
window). Therefore, the 
TDSE describing H$_2$ within the fixed-nuclei approximation was solved 
for 25 different values of $R$
(in between 1.0\,$a_0$ and 2.2\,$a_0$ with a step size of 0.05\,$a_0$). This 
was repeated for 15 different values of peak intensities 
(in a range from $2 \times 10^{13}$W/cm$^2$ to $1.3 \times 10^{14}$W/cm$^2$). 

Figure~\ref{fig:Ion_H2D2_All_CCgxxa}\,a (left panel) shows the obtained 
results for a parallel orientation of the molecular axis with respect 
to the polarization vector and 40-cycle cos$^2$-shaped (FWHM of about 
20\,fs) laser 
pulses. Two main features may be observed. First, the ionization yield 
increases with $R$. Second, pronounced structures are visible. The 
importance of both effects decreases with intensity. The increase 
with $R$ was first predicted in the quasistatic regime \cite{sfm:saen00c}. 
Its occurrence at 400\,nm shows that even for this wavelength 
clear strong-field phenomena are observable beyond a pure multiphoton 
picture. On the other hand, the observed structures are due to classical 
multiphoton phenomena (channel closings and REMPI).  

Some pronounced REMPI peaks are visible that should be compared  
with their predicted positions in Fig.~\ref{fig:H2_400nm_fig_3}. 
The peaks at $R=1.2\,a_0$ and $R=1.4\,a_0$ for 
a laser peak intensity of 
$2 \times 10^{13}$W/cm$^2$ can thus be assigned to (5+1) REMPI through 
the $2\,\Sigma_u^+$ and $3-4\,\Sigma_u^+$ states, respectively. 
The position of the latter peak changes with the peak intensity 
almost in the expected way as has also been demonstrated in 
Fig.~\ref{fig:Ion_H2R-aS_C-gxxa}\,a. On the other hand, the position and
amplitude of the REMPI peak arising from the $2\,\Sigma_u^+$ resonant state 
cannot easily be understood. For a peak intensity  of $10^{14}$W/cm$^2$ 
this REMPI peak is located at $R=1.6\,a_0$ instead of the expected value of 
$1.5\,a_0$. Interestingly, the amplitude of the 
peak that is very large at small intensities is becoming very small for higher
intensities, as one may expect when going from the multiphoton in the
direction of the quasistatic regime. 
For intensities between $4 \times 10^{13}$W/cm$^2$ and 
$6 \times 10^{13}$W/cm$^2$ one can observe something similar to a splitting of 
the REMPI peak into two peaks. This behavior can evidently not be explained 
using Fig.~\ref{fig:H2_400nm_fig_3}. It is also difficult to explain
the pronounced peak located at $R=1.4\,a_0$ for the highest laser peak 
intensity ($1.3 \times 10^{14}$W/cm$^2$). According to 
Fig.~\ref{fig:H2_400nm_fig_3} no peak should occur for these values of 
$R$ and $I$, since it lies in between the expected positions of the 
$1\,\Sigma_u^+$ and the $2\,\Sigma_u^+$ REMPI peaks. This may indicate 
some field-induced coupling of these states and thus a clear strong-field 
phenomenon. It is also interesting to note that the channel closings 
indicating the transitions from 5- to 6-photon ionization and from 6- to 
7-photon ionization are visible, but not very pronounced. Furthermore, 
due to the REMPI peaks the channel thresholds are sometimes difficult to 
identify in the shown ionization yields.  

In order to consider the influence of vibrational motion onto the 
strong-field ionization yields, the results obtained for a fixed 
nuclear orientation are weighted with the probability density 
of the ground vibrational state [see integrand of Eq.~(\ref{eq:Yionvibr})]. 
The corresponding result for H$_2$ is shown in 
the middle panel of Fig.~\ref{fig:Ion_H2D2_All_CCgxxa}. Only in the 
case of the highest laser peak intensity 
considered in this work, the maximum of the weighted ionization yield agrees 
with the maximum of the vibrational wave function. At slightly lower 
intensities (until about $8 \times 10^{13}$W/cm$^2$) the weighted ionization 
yield is largest for smaller values of $R$, while for even lower 
intensities the REMPI peaks due to the $3\,\Sigma_u^+$ and 
$4\,\Sigma_u^+$ states determine the maximum of the weighted ionization yield. 
At the lowest laser peak intensity considered ($2 \times 10^{13}$W/cm$^2$) 
the highest weighted ionization yield is found at around $R=1.7\,a_0$. This 
maximum should be due to an opening of the 5-photon regime and may be 
further increased by (4+1) REMPI processes. At this intensity one notices 
also a very slow decrease of the weighted ionization yield for $R$ values 
above $2\,a_0$, despite the fact that the vibrational wave function has 
a very small amplitude. The reason is the already discussed $3\,\Sigma_g^+$ 
REMPI peak (Fig.~\ref{fig:Ion_H2R-aS_C-gxxa}\,b).  

In the case of D$_2$ (left panel of Fig.~\ref{fig:Ion_H2D2_All_CCgxxa}\,a) 
the narrower vibrational distribution is, however, sufficient to dominate 
over the resonant effect. In this case the weighted ionization yield 
decreases rather pronouncedly for internuclear separations larger than 
$2.0\,a_0$. Close to the minimum of the electronic potential curve at 
about $1.4\,a_0$ the weighted ionization yields are, however, very similar 
for H$_2$ and D$_2$. Also for D$_2$ the ionization yield peaks only 
for the highest intensity considered here at the maximum of the 
vibrational wavefunction. Consequently, the Franck-Condon approximation 
would not describe the vibrational distribution of the formed H$_2^+$ 
ions properly. A proper calculation of these distributions has to include 
the effects of channel closings and REMPI, but also of the general increase 
of the ion yield as a function of internuclear separation. Neither a pure 
multiphoton nor quasistatic prediction is thus sufficient.

%%%%%%%%%%%%%%%%%%%%%%%%%%%%%%%%%%%%%%%%%%%%%%%%%%%%%%%%%%%%%%%%%%%%%%%%%%%%%
\subsection{Orientational dependence}
%%%%%%%%%%%%%%%%%%%%%%%%%%%%%%%%%%%%%%%%%%%%%%%%%%%%%%%%%%%%%%%%%%%%%%%%%%%%%

%
%%%%%%%%%%%%%%%%%%%%%%%%%%%%%%%%%%%%%%%%%%%%%%%%%%%%%%%%%%%%%%%%%%%%%%%%%%%%%
\begin{figure*}[!tbp]
\begin{center}
\includegraphics[width=12.0cm]{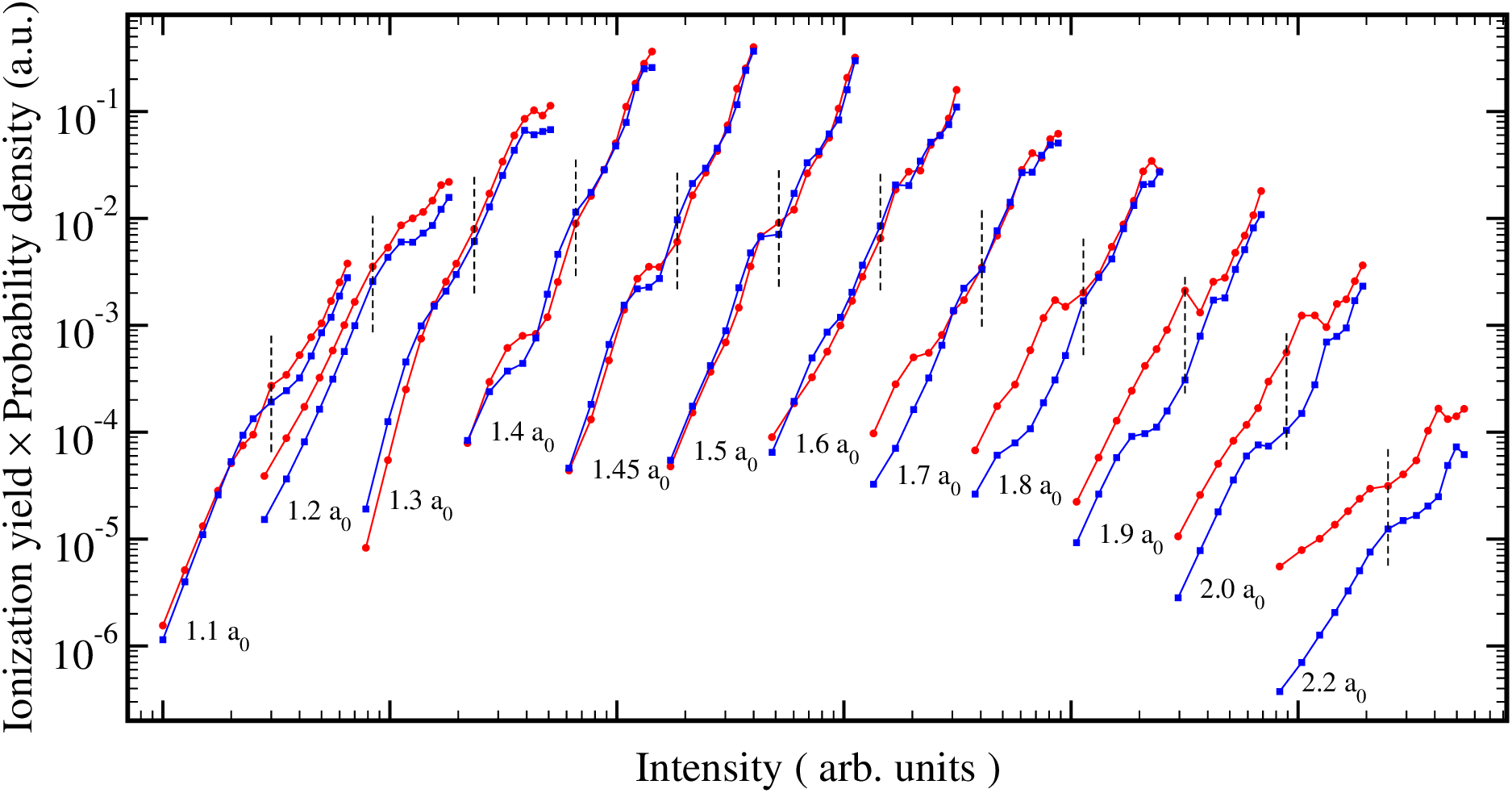}%
\caption{\label{fig:H2_400nm_fig_8} 
Comparison of ionization yields (in a 20\,fs laser pulse) for different 
internuclear distances $R$ 
(weighted with the the probability density of the ground vibrational 
state of H$_2$ at this $R$) for a parallel (red
curves) and a perpendicular (blue curves) oriented H$_2$ molecule. The 
$R$ values are specified explicitly in the figure. 
The vertical dashed lines indicate the position 
of a peak intensity $6 \times 10^{13}$W/cm$^2$ on the intensity axis 
for every corresponding pair of curves.} 
\end{center}
\end{figure*}
%%%%%%%%%%%%%%%%%%%%%%%%%%%%%%%%%%%%%%%%%%%%%%%%%%%%%%%%%%%%%%%%%%%%%%%%%%%%%
%

For a perpendicular orientation of the molecule with respect to the 
field (Fig.~\ref{fig:Ion_H2D2_All_CCgxxa}\,b) the $R$-dependent ionization 
yield (left panel) looks on the first glance surprisingly similar to the one 
for the parallel orientation. For the lowest shown intensities the spectra 
comprise very pronounced peaks at about $R=1.25\,a_0$ and $R=1.4\,a_0$ 
that can be assigned to (5+1) REMPI through the $2\,\Pi_u$ and the 
$3-4\,\Pi_u$ states, respectively. It is a peculiarity of H$_2$ that 
already the lowest lying excited states of $^1\Sigma_u$ and $^1\Pi_u$ 
symmetry and thus REMPI peaks through those states lie energetically very 
close together. In contrast to the results for parallel 
orientation a third peak at $R=1.55\,a_0$ is, however, also well 
resolved. From Fig.~\ref{fig:H2_400nm_fig_3} it appears very likely 
that this peak stems from a superposition of (5+1) REMPI processes 
through the higher excited $\Pi_u$ states. Especially at lower intensities 
one notices furthermore that the ionization yield does not increase that 
evidently for large $R$ values than it does for a parallel orientation. 
The threshold between 5- and 6-photon ionization is rather well 
resolved and appears for the different laser peak intensities more or less 
at the expected $R$ values (Fig.~\ref{fig:H2_400nm_fig_3}\,a). 
As a consequence of the smaller slope at large $R$ the weighted ionization 
yield decays for larger $R$ values much faster for the perpendicular than 
for the parallel orientation for both H$_2$ (middle panel of 
Fig.~\ref{fig:Ion_H2D2_All_CCgxxa}\,b) and D$_2$ (right panel).  

Figure \ref{fig:Ion_H2D2_All_CCgxxa}\,c shows finally the ionization yield 
obtained with the simple isotropic one-electron model potential given in 
Eq.~(\ref{eq:Vatom}). The agreement of the $R$-dependent ionization yields  
obtained with this model and the full molecular two-electron calculation 
are surprisingly good, especially with the results obtained for the 
parallel orientation. As in the latter case, the atomic model gives  
a shifted threshold between the 5-and 6-photon regimes compared to the 
prediction according to Fig.~\ref{fig:H2_400nm_fig_3}. The atomic 
model yields also a rather pronounced increase in ionization for large 
$R$ values, especially for low laser peak intensities as was also found for 
the parallel orientation. A closer look reveals, however, that for small 
$R$ separations and especially for the first REMPI peak the atomic 
model agrees slightly better with the molecular results obtained for the 
perpendicular orientation. The 2nd REMPI peak defines somehow the transition 
line. For smaller $R$ values the atomic model agrees better with the 
perpendicular results, while starting with the 2nd REMPI peak the ionization 
yields obtained for the atomic model and the molecular one for a parallel 
orientation agree better with each other. The main difference to the molecular 
calculations is the position of the 3rd REMPI peak that for the lowest shown 
intensity lies so close to the 2nd one, that it appears in the $R$-dependent 
ionization yield as a shoulder. For a laser peak intensity of 
$2.5 \times 10^{13}$W/cm$^2$ the 3rd REMPI peak is shifted more than the 
2nd one and is thus visible as a well separated peak. However, for 
higher intensities it is less well resolved due to its low probability. 
Despite the overall good agreement of the results for the atomic model 
with the full molecular calculations (on a logarithmic scale!), the weighted 
ionization yields still reveal differences. For example, the maximum of the 
weighted ionization yields for H$_2$ and D$_2$ and the largest laser peak 
intensities is shifted to slightly smaller $R$ values than is found for the 
full molecular calculations. 

A further important laser parameter is the pulse duration. Its influence 
is demonstrated in Figs.~\ref{fig:Ion_H2D2_All_CCcxxa} and 
\ref{fig:Ion_H2D2_All_CC5xxa} that show the corresponding results 
for 20- and 10-cycle pulses (FWHM of 10 and 5\,fs), respectively. The 
increased laser bandwidth leads to spectra that show much less details 
compared to the relatively long 40-cycle pulse. 
The ionization yields for the 10-cycle pulse show almost no evidence of REMPI
peaks. The curves are fairly smooth and the remaining structures can be 
explained by the closing and opening of $N$-photon ionization channels.  

Because of the different positions of the REMPI peaks for parallel or 
perpendicular orientations the ratio of parallel to perpendicular 
ionization yields may substantially change for a small variation of $R$. 
This effect is demonstrated in Fig.~\ref{fig:H2_400nm_fig_8} in which 
ionization yields (multiplied with the probability density of the ground 
vibrational state) for parallel and perpendicular  
oriented H$_2$ molecule are compared for different internuclear distances. 
A log-log scale is used and the pairs of curves (parallel and perpendicular 
orientation for a given value of $R$) are shifted along the intensity axis 
for better readability. 
To guide the eye, the vertical dashed lines indicate the position of the 
peak intensity $6 \times 10^{13}$W/cm$^2$ on the intensity axis for every 
pair of curves.  The multiplication with the probability density was 
performed in order to emphasize the relative contributions of different
internuclear distances for the total ionization ratio between parallel 
and perpendicular orientations obtained after integration over $R$. 

As can be seen from Fig.~\ref{fig:H2_400nm_fig_8}, the ionization 
yields for parallel and perpendicular orientation are almost equal 
in the range $R=1.3-1.7\,a_0$ for a peak intensity of 
$6 \times 10^{13}$W/cm$^2$. Whereas the ionization yield for the
parallel orientation is larger than for the perpendicular one at
$R=1.3$ and $1.5\,a_0$, the opposite is found at $R=1.4, 1.45$, and 
$1.6\,a_0$. At smaller values of $R$ the parallel orientation is 
slightly easier ionized than the perpendicular one, while for $R$ 
values larger than $1.8\,a_0$ parallel oriented molecules are 
much easier ionized. This is a consequence of the slower decay of 
the ionization yield for a parallel orientation and for large $R$ 
values that was already discussed in the context of 
Figs.~\ref{fig:Ion_H2D2_All_CCgxxa}\,a and b. 

The key conclusion that can be drawn from Fig.~\ref{fig:H2_400nm_fig_8} 
is the need for systematic studies of the intensity and 
internuclear-separation dependencies of the ratio between the 
ionization yields for parallel or perpendicular orientation as they 
are performed in this work, since a calculation for a single laser peak 
intensity and internuclear separation $R$ can yield any possible 
result, i.\,e.\ the ratio between the ionization yields for parallel 
and perpendicular orientations may be found to be equal to 1, much 
smaller than 1, or much larger than 1. Depending on the choice of 
intensity and $R$ very different conclusions on the orientation 
dependence of the ionization yield of H$_2$ in strong laser fields 
would follow.

%%%%%%%%%%%%%%%%%%%%%%%%%%%%%%%%%%%%%%%%%%%%%%%%%%%%%%%%%%%%%%%%%%%%%%%%%%%%%
\subsection{\label{subsec:IntYion}Integrated ionization yields}
%%%%%%%%%%%%%%%%%%%%%%%%%%%%%%%%%%%%%%%%%%%%%%%%%%%%%%%%%%%%%%%%%%%%%%%%%%%%%

%
%%%%%%%%%%%%%%%%%%%%%%%%%%%%%%%%%%%%%%%%%%%%%%%%%%%%%%%%%%%%%%%%%%%%%%%%%%%%%
\begin{figure*}[!tbp]
\centering
\subfigure[10-cycle cos$^2$-shaped (5\,fs) laser pulse]{
\includegraphics[width=16.0cm]{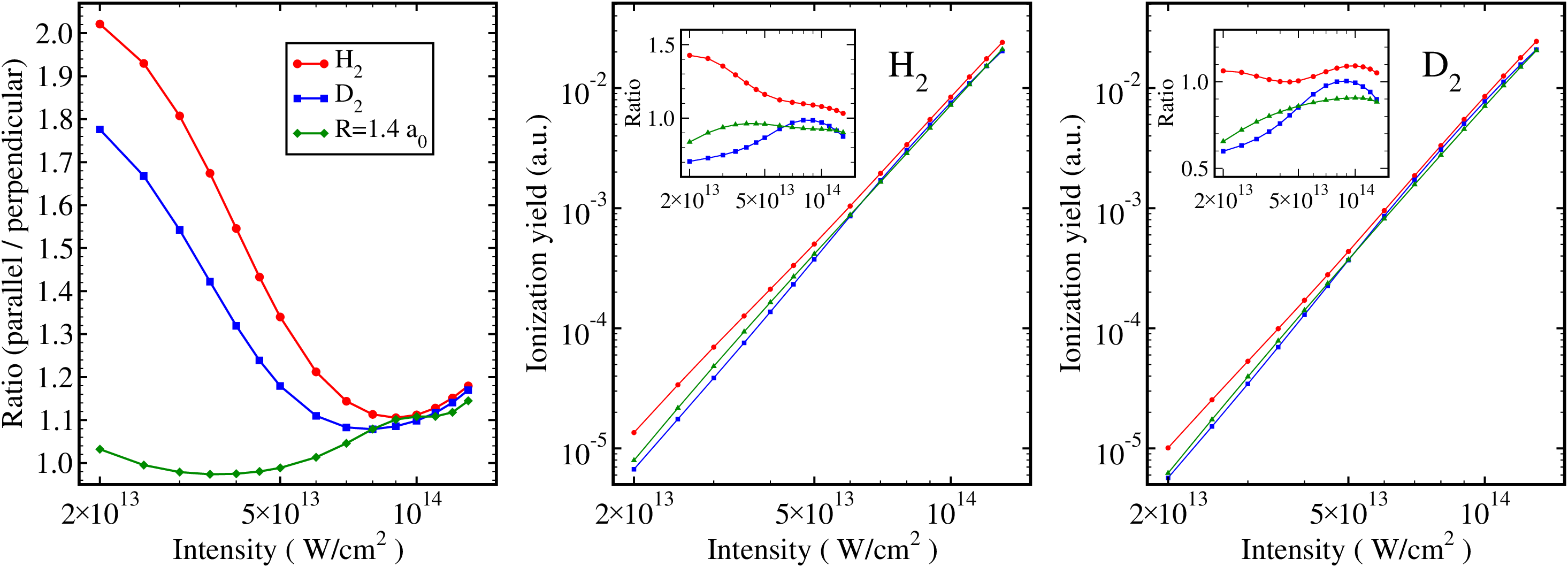}%
}\\
\subfigure[20-cycle cos$^2$-shaped (10\,fs) laser pulse]{
\includegraphics[width=16.0cm]{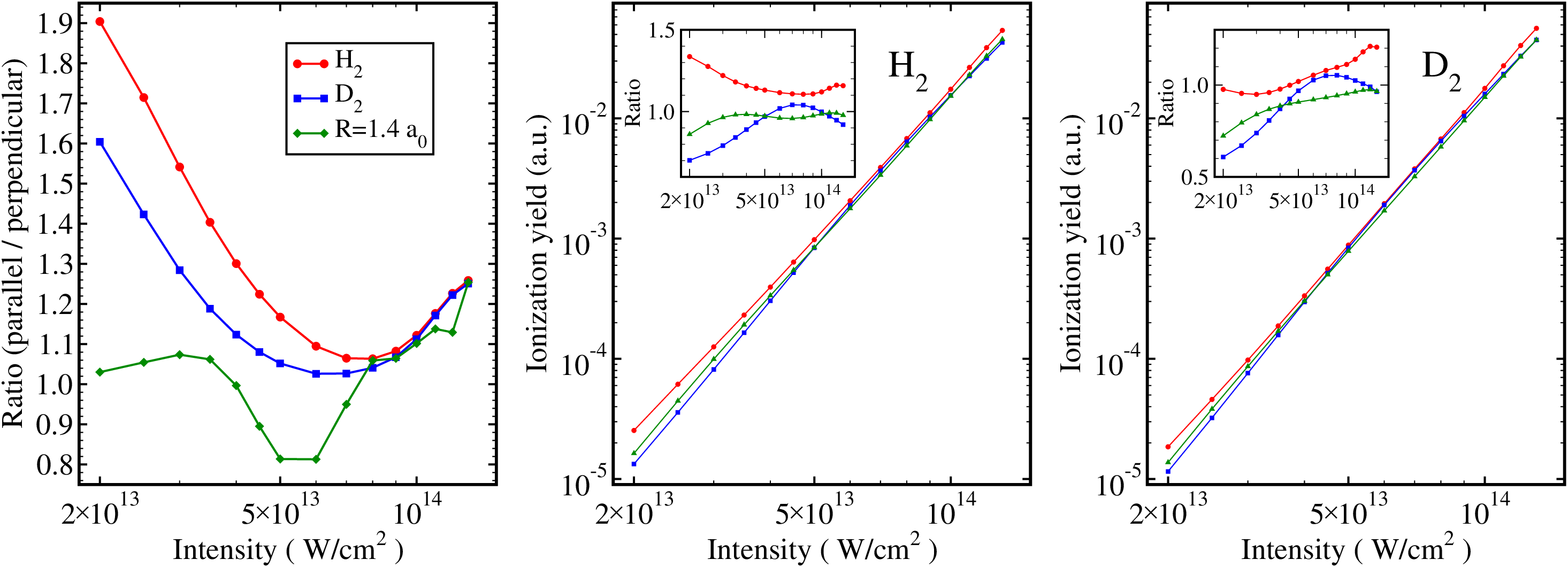}%
}\\
\subfigure[40-cycle cos$^2$-shaped (20\,fs) laser pulse]{
\includegraphics[width=16.0cm]{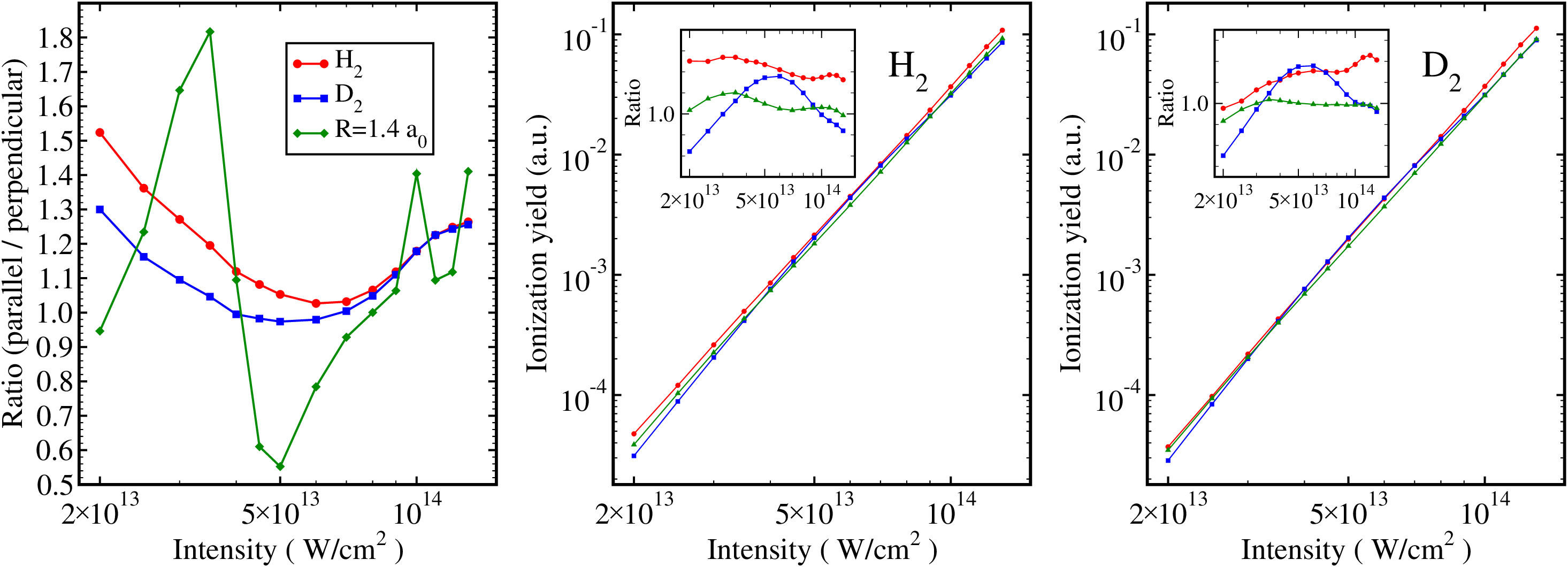}%
}%
\caption{\label{fig:IntY_H2D2_All} 
Final ionization yields (integrated over the internuclear separation $R$) 
as a function of the laser peak intensity for 
H$_2$ (middle panel) and D$_2$ (right panel) and a parallel (red circles) 
or a perpendicular (blue squares) orientation and a) 10-cycle, b) 20-cycle, 
or c) 40-cycle laser pulses. Also shown are the results 
obtained with the isotropic one-electron model potential (green triangles). 
The inserts show the ratio of the ionization yields to 
the fit function in Eq.~(\ref{eq:IntY_scal}). The resulting ratio 
of parallel to perpendicular ionization yields is shown in the left panel 
in which also the corresponding ratio obtained for a fixed internuclear 
separation ($R=1.4\,a_0$) is plotted.}
\end{figure*}
%%%%%%%%%%%%%%%%%%%%%%%%%%%%%%%%%%%%%%%%%%%%%%%%%%%%%%%%%%%%%%%%%%%%%%%%%%%%%
%

The $R$-independent ionization yields for H$_2$ and D$_2$ molecules 
(in their vibrational ground states) are finally obtained by an 
integration of the weighted $R$-dependent ionization yields over $R$ 
(Eq.\,(\ref{eq:Yionvibr}) in Sec.~\ref{subsec:Integr}). The results 
for parallel and perpendicular orientation are shown for the different 
pulse lengths in Fig.~\ref{fig:IntY_H2D2_All}. The curves obtained 
after $R$ integration are much less structured than the fixed-nuclei
ionization yields, as is evident from a comparison with 
Fig.~\ref{fig:H2_400nm_fig_8}). Clearly, the structures due to REMPI 
processes are smoothed out by the integration over $R$. As a consequence, 
the curves look almost like straight lines on the used log-log scale. 

In fact, it turns out that the linear dependence on the logarithmic scale 
is well described, if the yield is fitted with the function%
%---------------
\begin{equation}
       Y_s(I,T) = \Omega\, T\, (I/I_0)^{k_s} 
\label{eq:IntY_scal}
\end{equation}
%---------------
where $I$ is the peak intensity, $I_0 = 3.5094452 \times 10^{16}\,$ W/cm$^2$ is 
the atomic unit of intensity, and $T$ is the FWHM duration of the pulse in
atomic units. For the fit parameters the values $\Omega = 1.55 \cdot 10^6$ 
and $k_s = 4.17$ are found. The obtained value of $k_s$ indicates a 
non-perturbative behavior, since according to 
Fig.~\ref{fig:H2_400nm_fig_3}\,a one would expect mostly 6-photon 
ionization to occur and thus $k_s$ should be close to 6. The included 
dependence on the pulse duration $T$ allows to compare the results obtained 
for different pulse lengths. A linear dependence on $T$ should be found, 
if a rate concept is applicable.  

Dividing the ionization yields by the fit function (\ref{eq:IntY_scal}) 
allows a direct comparison of the parallel, perpendicular, and atomic 
model potential results on a linear scale. They are shown in the inserts 
of Fig.~\ref{fig:IntY_H2D2_All}. If the vibrational ground state of H$_2$ 
is considered (middle panel of Fig.~\ref{fig:IntY_H2D2_All}), the scaled 
atomic-model results are closest to 1 and thus most accurately described 
by the fit function. The scaled yield for parallel orientation decreases 
from a value of 1.5 to about 1.0 for the 10-cycle pulse, but shows an 
increasing behavior for higher intensities in the case of the 20-cycle 
pulse. The smallest intensity dependence is found for the longest pulse 
considered in this work where the scaled yield varies only between about 
1.25 and 1.15. Interestingly, the scaled yield for perpendicular 
orientation shows almost the opposite behavior. The most pronounced 
intensity dependence is found for the longest pulse. Furthermore, 
the scaled yield increases for low intensities as a function of intensity. 
As a consequence, the scaled yields for parallel and perpendicular 
orientation first approach each other before they separate again for even 
larger intensities. 

Using the same fit function for scaling the D$_2$ yields one notices that 
the yields for parallel orientation are now almost flat (shortest pulse) 
or increase with intensity. Since the vibrational density distribution 
of D$_2$ is more localized around $R_0$, one can conclude that such higher  
ratio for H$_2$ can be due to contributions to the ionization from either 
small or large internuclear distances $R$. The comparison of the middle and 
the right panels of Fig.~\ref{fig:Ion_H2D2_All_CC5xxa}\,a shows that the 
effect stems from the enhanced ionization at $R>1.6\,a_0$. Also the scaled 
yields for perpendicular orientation or the atomic model potential show 
a larger increase with intensity for D$_2$ compared to H$_2$, 
although this effect is a little bit less pronounced. This indicates that the 
ionization yield of D$_2$ possesses a slightly larger slope than the one 
of H$_2$, a rather unexpected (though small) isotope effect.  

Finally, the ratio of the ionization yields for parallel to perpendicular 
orientation of the molecular axis as a function of the peak intensity is also 
shown in  Fig.~\ref{fig:IntY_H2D2_All}. 
For the 10-cycle pulse and H$_2$ the ratio is about 2 for the peak intensity  
$2 \times 10^{13}$W/cm$^2$ and decreases smoothly to about 1.1 at
$9 \times 10^{13}$W/cm$^2$, before it increases to 1.18 at 
$1.3 \times 10^{14}$W/cm$^2$. The occurrence of a minimum is due to 
the maximum found for scaled yield in the case of the perpendicular 
orientation, as was discussed in the context of the inserts in the 
middle and right panels of Fig.~\ref{fig:IntY_H2D2_All}. 
Increasing the pulse duration does not 
change the behavior in a qualitative fashion, but the ratio found at 
small intensities decreases with increasing pulse length. At the same 
time, the increase at the highest intensities is more pronounced, but 
this increase is smaller than the decrease seen for the low intensities. 
In the case of the 40-cycle pulse the ratio starts at about 1.5, decreases 
to almost 1.0 and increases to 1.26. The turning point shifts also 
to slightly lower intensities for longer pulses. The intensity dependence 
of the ratios for D$_2$ show a similar behavior as was found for H$_2$. 
However, at small intensities the ratio is clearly smaller than for 
H$_2$, i.\,e.\ the anisotropy of the ionization yield is less pronounced. 
For high intensities the ratios found for H$_2$ and D$_2$ agree on the 
other hand almost perfectly with each other. An isotope effect occurs 
thus only for low intensities. 

The importance of the inclusion of nuclear motion is evident from the 
ratio of parallel to perpendicular ionization yields obtained for 
a fixed internuclear separation ($R=1.4\,a_0$) that is also shown in 
Fig.~\ref{fig:IntY_H2D2_All}.  In the case of a 10-cycle pulse the 
ratio is also smooth, but increases with intensity. The pronounced 
decrease found for the $R$-integrated ratio at low intensities is 
thus completely absent. Interestingly, the agreement with the ratio 
found for H$_2$ is very good for high intensities for which also the 
H$_2$ and D$_2$ ratios agreed well with each other. This indicates 
that for high intensities the ratio is less sensitive to $R$. The 
reason is the less pronounced $R$ dependence of the ionization yields 
for high intensities that was found in general and discussed in the 
context of the weighted $R$-dependent ionization yields in 
Figs.~\ref{fig:Ion_H2D2_All_CCgxxa} to 
\ref{fig:Ion_H2D2_All_CC5xxa}. One of the consequences of this reduced 
$R$ dependence was, e.\,g., that the maximum of the ionization yield 
was more or less found for the $R$ value at which the vibrational 
density had its maximum. For longer pulses pronounced 
intensity-dependent maxima and minima become visible. As a consequence, 
the ratio found for a single $R$ value differs clearly from the 
$R$-integrated results. For example, in the case of a 40-cycle pulse 
and $R=1.4\,a_0$ the ratio decreases to about 0.5 
at a laser peak intensity of $5 \times 10^{13}$W/cm$^2$ which means that 
perpendicular oriented H$_2$ ionizes much better than parallel oriented 
one. This is in complete contrast to the $R$-integrated results for which 
the perpendicular orientation never ionizes faster than parallel 
oriented molecules in the considered range of laser peak intensities.

%%%%%%%%%%%%%%%%%%%%%%%%%%%%%%%%%%%%%%%%%%%%%%%%%%%%%%%%%%%%%%%%%%%%%%%%%%%%%
%%%%%%%%%%%%%%%%%%%%%%%%%%%%%%%%%%%%%%%%%%%%%%%%%%%%%%%%%%%%%%%%%%%%%%%%%%%%%
\section{\label{sec:Concl}Conclusion}
%%%%%%%%%%%%%%%%%%%%%%%%%%%%%%%%%%%%%%%%%%%%%%%%%%%%%%%%%%%%%%%%%%%%%%%%%%%%%
%%%%%%%%%%%%%%%%%%%%%%%%%%%%%%%%%%%%%%%%%%%%%%%%%%%%%%%%%%%%%%%%%%%%%%%%%%%%%

We have performed an extensive numerical study of 
intense-field ionization of molecular hydrogen and deuterium numerically
integrating the full-dimensional two-electron Schr\"odinger equation in
the non-relativistic, fixed-nuclei, and dipole approximation.
The presented results are obtained for three different durations 
($5, 10, 20\,$fs) of ultrashort frequency-doubled Ti:sapphire laser 
pulses (400\,nm) for both parallel and perpendicular orientations of the 
molecular axis with respect to the laser field. Calculations are performed 
for 15 different intensities (in a range from $2\times10^{13}\,$W\,cm$^{-2}$ 
to  $1.3\times10^{14}\,$W\,cm$^{-2}$)
and 25 different internuclear separations (in a range from $1.0\,a_0$ to
$2.2\,a_0$) which results in 375 data points for each orientation. The same
series of calculations was performed employing an isotropic one-electron 
model potential in order to study the influence of molecular anisotropy or 
the one of the two electrons. 

By analyzing the dependence of the fixed-nuclei ionization yields on the peak 
intensity and the internuclear separation, we assign the observed peaks to 
REMPI or closings of $N$-photon ionization channels and study field-induced 
shifts of resonant electronic states. 

A key feature of the present work is the calculation of total ionization yields
of  molecular hydrogen and deuterium in their ground vibrational states
by an integration of the fixed-nuclei ionization yields multiplied with 
the corresponding vibrational probability density  over the internuclear 
separation. The subsequent analysis of the ionization anisotropy reveals 
a smooth dependence on the laser peak intensity and pulse duration. 
The obtained ratios of the parallel to perpendicular total ionization yields 
vary in the investigated intensity and pulse duration ranges in between 
1 and 2. Whereas for high intensities the ionization anisotropy of H$_2$ 
and D$_2$ is almost identical, the ratio for H$_2$ is always larger than 
the one for D$_2$ in the case of small intensities, although the difference 
does not exceed 20\,\%. 

The importance of the integration over the internuclear separation for the 
interpretation of experiments is demonstrated by comparing the obtained 
parallel to perpendicular ratios with those calculated using the 
fixed-nuclei results at the equilibrium distance.
For example, there is a wide range of intensities where at the 
equilibrium distance the ionization for perpendicular-oriented molecule is 
larger than for parallel-oriented ones. Besides the rather different 
qualitative dependence on the peak intensity, also the dependence on 
the pulse duration is different.

The results of this work do not only provide an interesting 
insight into the strong-field behavior of molecules and the influence 
of channel closings, REMPI, field-induced shifts of energy positions, 
pulse length and intensity, vibrational motion, isotope effects, and 
orientation, but will hopefully also stimulate experimental efforts 
to measure the anisotropy of the ionization yield of H$_2$ (or D$_2$) 
in the interesting regime that is neither well described by the 
multiphoton nor the quasistatic approximations. Furthermore, the 
results should serve as benchmarks for other theoretical approaches 
and simplified models.

%%%%%%%%%%%%%%%%%%%%%%%%%%%%%%%%%%%%%%%%%%%%%%%%%%%%%%%%%%%%%%%%%%%%%%%%%%%%%
%%%%%%%%%%%%%%%%%%%%%%%%%%%%%%%%%%%%%%%%%%%%%%%%%%%%%%%%%%%%%%%%%%%%%%%%%%%%%
\section*{Acknowledgments}
%%%%%%%%%%%%%%%%%%%%%%%%%%%%%%%%%%%%%%%%%%%%%%%%%%%%%%%%%%%%%%%%%%%%%%%%%%%%%
%%%%%%%%%%%%%%%%%%%%%%%%%%%%%%%%%%%%%%%%%%%%%%%%%%%%%%%%%%%%%%%%%%%%%%%%%%%%%
This work was supported by the Deutsche
Forschungsgemeinschaft (Sa936/2) and COST action CM0702. AS is grateful
to the Stifterverband f\"ur die Deutsche Wissenschaft (Program
``Forschungsdozenturen'') and the Fonds der Chemischen Industrie
for financial support.

%%%%%%%%%%%%%%%%%%%%%%%%%%%%%%%%%%%%%%%%%%%%%%%%%%%%%%%%%%%%%%%%%%%%%%%%%%%%%

\bibliographystyle{apsrev}
%\bibliography{sfa,sfm,bsp,dia,gen,anti}

%-----------------------------------------------------------------

\end{document}